\documentclass[fleqn,10pt]{wlscirep}
\usepackage[utf8]{inputenc}
\usepackage[T1]{fontenc}
\usepackage{marvosym}
\usepackage{authblk}
\usepackage{xcolor}
\usepackage{soul}

\usepackage{diagbox}
\usepackage{amsmath}
\newcommand{\argmin}{\mathop{\mathrm{argmin}}\limits}
\usepackage{subcaption}
\usepackage{amssymb} 

\newcommand{\yes}{\checkmark} 
\newcommand{\no}{$\times$}    

\usepackage{tabulary}
\usepackage{float}

\title{Multi-Parameter Molecular MRI Quantification using Physics-Informed Self-Supervised Learning}

\author[1]{Alex Finkelstein}
\author[1]{Nikita Vladimirov}
\author[2,3]{Moritz Zaiss}
\author[1,4*]{Or Perlman}
\affil[1]{Department of Biomedical Engineering, Tel Aviv University, Tel Aviv, Israel}
\affil[2]{Institute of Neuroradiology, University Hospital Erlangen, Friedrich-Alexander-Universität Erlangen-Nürnberg (FAU), Erlangen, Germany}
\affil[3]{Department of Artificial Intelligence in Biomedical Engineering, Friedrich-Alexander-Universität Erlangen-Nürnberg (FAU), Erlangen, Germany}
\affil[4]{Sagol School of Neuroscience, Tel Aviv University, Tel Aviv, Israel}

\affil[*]{Correspondence: orperlman@tauex.tau.ac.il}
\vspace{2pt}

\begin{abstract}
Biophysical model fitting plays a key role in obtaining quantitative parameters from physiological signals and images. However, the model complexity for molecular magnetic resonance imaging (MRI) often translates into excessive computation time, which makes clinical use impractical. Here, we present a generic computational approach for solving the parameter extraction inverse problem posed by ordinary differential equation (ODE) modeling coupled with experimental measurement of the system dynamics. This is achieved by formulating a numerical ODE solver to function as a step-wise analytical one, thereby making it compatible with automatic differentiation-based optimization. This enables efficient gradient-based model fitting, and provides a new approach to parameter quantification based on self-supervised learning from a single data observation. The neural-network-based train-by-fit pipeline was used to quantify semisolid magnetization transfer (MT) and chemical exchange saturation transfer (CEST) amide proton exchange parameters in the human brain, in an in-vivo molecular MRI study (n=4). The entire pipeline of the first whole brain quantification was completed in 18.3$\pm$8.3 minutes. 
Reusing the single-subject-trained network for inference in new subjects took 1.0$\pm$0.2 s, to provide results in agreement with literature values and scan-specific fit results.

\end{abstract}
\begin{document}

\flushbottom
\maketitle
\begin{raggedleft}
Published in \emph{Communications Physics} at \url{https://doi.org/10.1038/s42005-025-02063-8}. This version includes the main text and the Supplementary Information in a single document.
\end{raggedleft}
\thispagestyle{empty}

\section*{Introduction}
Magnetic resonance imaging (MRI) plays a central role in clinical diagnosis and neuroscience. This modality is highly versatile and can be selectively programmed to generate a large number of image contrasts\cite{Bernstein2004}, each sensitive to certain biophysical parameters of the tissue. In recent years, there has been extensive research into developing quantitative MRI (qMRI) methods that can provide reproducible measurements of magnetic tissue properties (such as: T$_1$, T$_2$, and T$_2^*$), while being agnostic to the scan site and the exact acquisition protocol used\cite{Weiskopf2021}. Classical qMRI quantifies each biophysical property separately\cite{radunsky2024comprehensive}, using repeated acquisition and gradual variation of a single acquisition parameter under steady state conditions. This is followed by fitting the model to an analytical solution of magnetization vector dynamics\cite{taylor2016t1}. 

The exceedingly long acquisition times associated with the classical quantification pipeline have motivated the development of magnetic resonance fingerprinting (MRF)\cite{Ma2013}, an alternative paradigm for the joint extraction of multiple tissue parameter maps from a single pseudorandom pulse sequence. Since MRF data are acquired under non-steady state conditions\cite{panda2017magnetic}, the corresponding magnetization vector can only be resolved numerically. This comes at the expense of the complexity of the inverse problem, namely finding tissue parameters that best reconstruct the signal according to the forward model of spin dynamics. Since model fitting under these conditions takes an impractically long time\cite{heo2019quantifying}, MRF is commonly solved by dictionary matching, where a large number of simulated signal trajectories are compared to experimentally measured data\cite{poorman2020magnetic}. Unfortunately, the size of the dictionary scales exponentially with the number of parameters (the "curse of dimensionality"\cite{hsieh2020magnetic}), which rapidly escalates the compute and memory demands of both generation and subsequent use of the dictionary for pattern matching-based inference.

Recently, various deep learning (DL)-based methods have been developed for replacing the lengthy dictionary matching with neural-network (NN)-based inference\cite{Cohen2018DRONE,Kim2020,Perlman2022_NatureBME_Apoptosis,Cohen2023}. While this approach greatly reduces the parameter quantification time, networks still need to be trained using a comprehensive dictionary of synthetic signals. Since dictionary generation may take days\cite{Perlman2022_NatureBME_Apoptosis}, it constitutes an obvious bottleneck for routine use of MRF, and reduces the possibilities for addressing a wide variety of clinical scenarios. Even with a faster generation, the transfer of synthetic data-trained NN to experimental data raises concerns about biased estimates (See Supplementary Note 2).

The complexity and time constraints associated with the MRF pipeline are drastically exacerbated for molecular imaging applications that involve a plurality of proton pools, such as chemical exchange saturation transfer (CEST) MRI\cite{perlman2020cest}. While CEST has demonstrated great potential for dozens of biomedical applications\cite{Jones2018, Zhou2019, vinogradov2023cest, bricco2023genetic, vladimirov2023molecular, wang2024creatine, rivlin2023metabolic}, some on the verge of entering clinical practice\cite{Zhou2022}, the inherently large number of tissue properties greatly complicate analysis\cite{Wu2016}. This has prompted considerable efforts to transition from CEST-weighted imaging to fully quantitative mapping of proton exchange parameters \cite{
ji2017progress,zaiss2018quesp, meissner2015quantitative, woessner2005numerical, Perlman2023_Review_w_Heo}. Early CEST quantification used the fitting of the classical numerical model (based on the underlying Bloch-McConnell equations) after irradiation at various saturation pulse powers ($B_1$)\cite{woessner2005numerical}. However, applying this approach in a pixelwise manner in-vivo is unrealistic because both the acquisition and reconstruction steps may require several hours. Later, faster approaches, such as quantification of the exchange by varying saturation power/time and Omega-plots \cite{Zaiss2018_QUESPrev, mcmahon2006quantifying, Meissner2015, Wu2015_OmegaPlot} still rely on steady-state (or close to steady state) conditions, and approximate analytical expressions of the signal as a function of the tissue parameters\cite{Zaiss2015_analytic3pool, Roeloffs2015}. Unfortunately, a closed-form analytical solution does not exist for most practical clinical CEST protocols, which utilize a train of off-resonant radiofrequency (RF) 
pulses saturating multiple interacting proton pools. Similarly to the quantification of water T$_1$ and T$_2$, incorporating the concepts of MRF into CEST studies provided new quantification capabilities\cite{Perlman2023_Review_w_Heo, Cohen2018CESTMRF, Zhou2018, heo2024unraveling, cohen2023cest} and subsequent biological insights, for example, in the detection of apoptosis after oncolytic virotherapy\cite{Perlman2022_NatureBME_Apoptosis}. However, in order to further push the boundaries of CEST MRF 
research and expedite its progress, the long dictionary generation associated with each new application needs to be replaced by a rapid and flexible approach that adequately models multiple proton pools under saturation pulse trains.

\begin{figure}[ht]
\centering
\includegraphics[trim={0 0 0 4.0cm},clip,width=\linewidth]{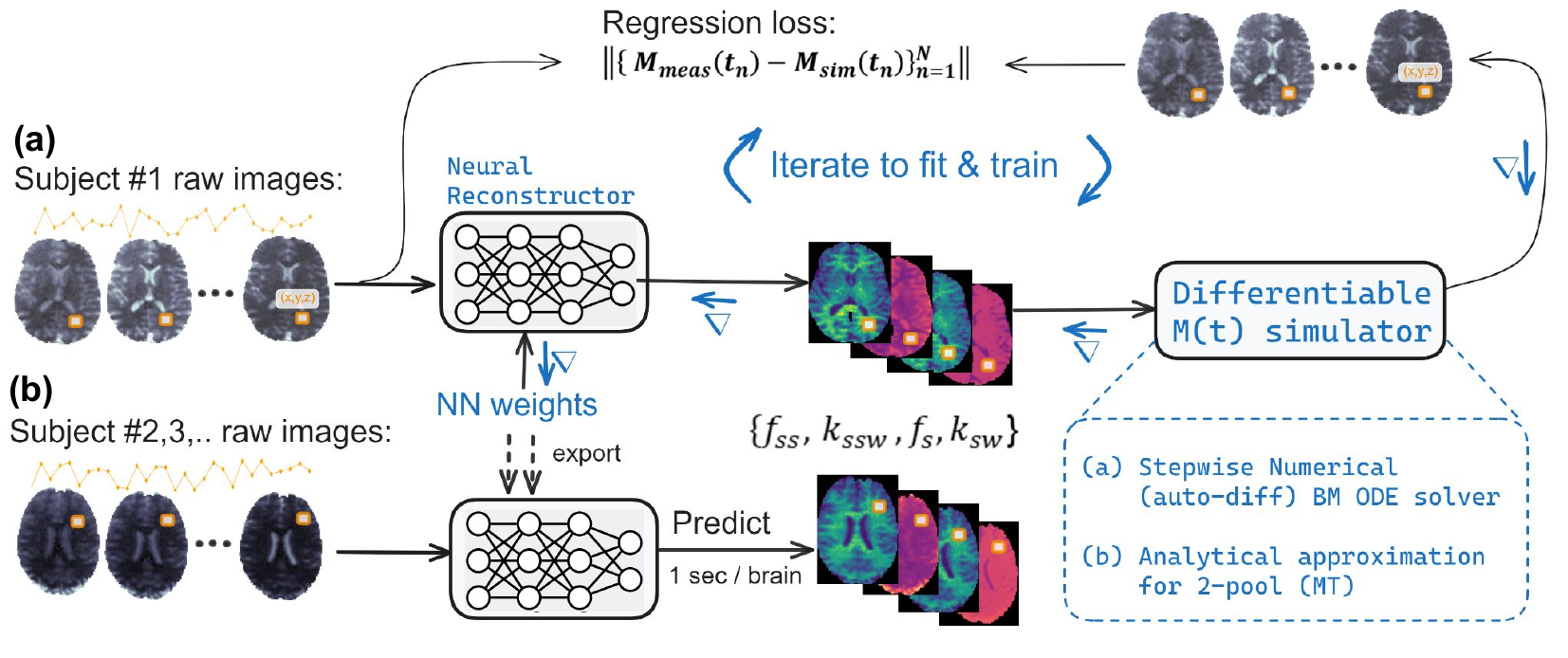}
\caption{Schematic representation of the core neural Bloch McConnell fitting (NBMF) pipeline. A  quantitative parameter reconstructor parameterized as a multi-layer perceptron (MLP) and a differentiable Bloch-McConnell simulator are serially connected into a single computational graph. Single-subject MRF data serves both as the input and as the regression target for the reconstructor-simulator circuit. The network convergence (a) provides the fitted exchange parameter maps for the examined subject as well as a trained NN reconstructor; the latter can be used to extract parameter maps for new subjects within seconds (b). The simulator can be realized using the exact numerical Bloch McConnell ODE solver or using analytical approximations when available (e.g., for 2-pool semisolid-MT quantification \cite{Roeloffs2015}). While not shown in the diagram, auxiliary per-voxel data such as T$_1$, T$_2$, B$_0$, and B$_1$ maps can be added as input to the neural reconstructor and the simulator. Furthermore, the pipeline main block can be serially repeated so that estimated semisolid MT volume fraction (f$_{ss}$) and proton exchange rate (k$_{ssw}$) maps inferred at the first stage are joined to the raw data used in a second reconstructor aimed to quantify the amide proton exchange parameters (f$_s$, k$_{sw}$).
}
\label{fig:nbmf_overview}
\end{figure}

Here, we describe a physics-based deep learning framework for rapid model fitting of the human brain tissue proton spin properties. While this approach is applicable for quantifying a variety of MRI parameters, we focus on a challenging CEST imaging scenario, involving multiple proton pools, a saturation pulse train, and non-steady-state MRF acquisition. The computational pipeline (Fig. \ref{fig:nbmf_overview}) combines a spin physics simulator and a NN-based quantitative parameter reconstructor in a fully auto-differentiable manner\cite{jax2018github}. Our system effectively solves and inverts the Bloch-McConnell ordinary differential equations (ODEs), which govern the  multi-pool exchange, saturation, and relaxation dynamics of molecular MRI. Hence, we refer to this approach as "neural Bloch McConnell fitting" (NBMF). Importantly, the network can be be trained in a self-supervised manner, directly on the single-subject data of interest (inspired by related work on test-time-\cite{Sun2020}, internal-\cite{Shocher2018}, and zero-shot-\cite{Ulyanov2020,Shocher2018,Yaman2022} learning). This circumvents the need for prior curation of a large training dataset, which is often inaccessible, especially for molecular MRI. 

\section*{Results}

\subsection*{In-vitro CEST quantification}
\begin{figure}[ht]
\centering
\includegraphics[trim={0 0 0 0cm},clip,width=\linewidth]{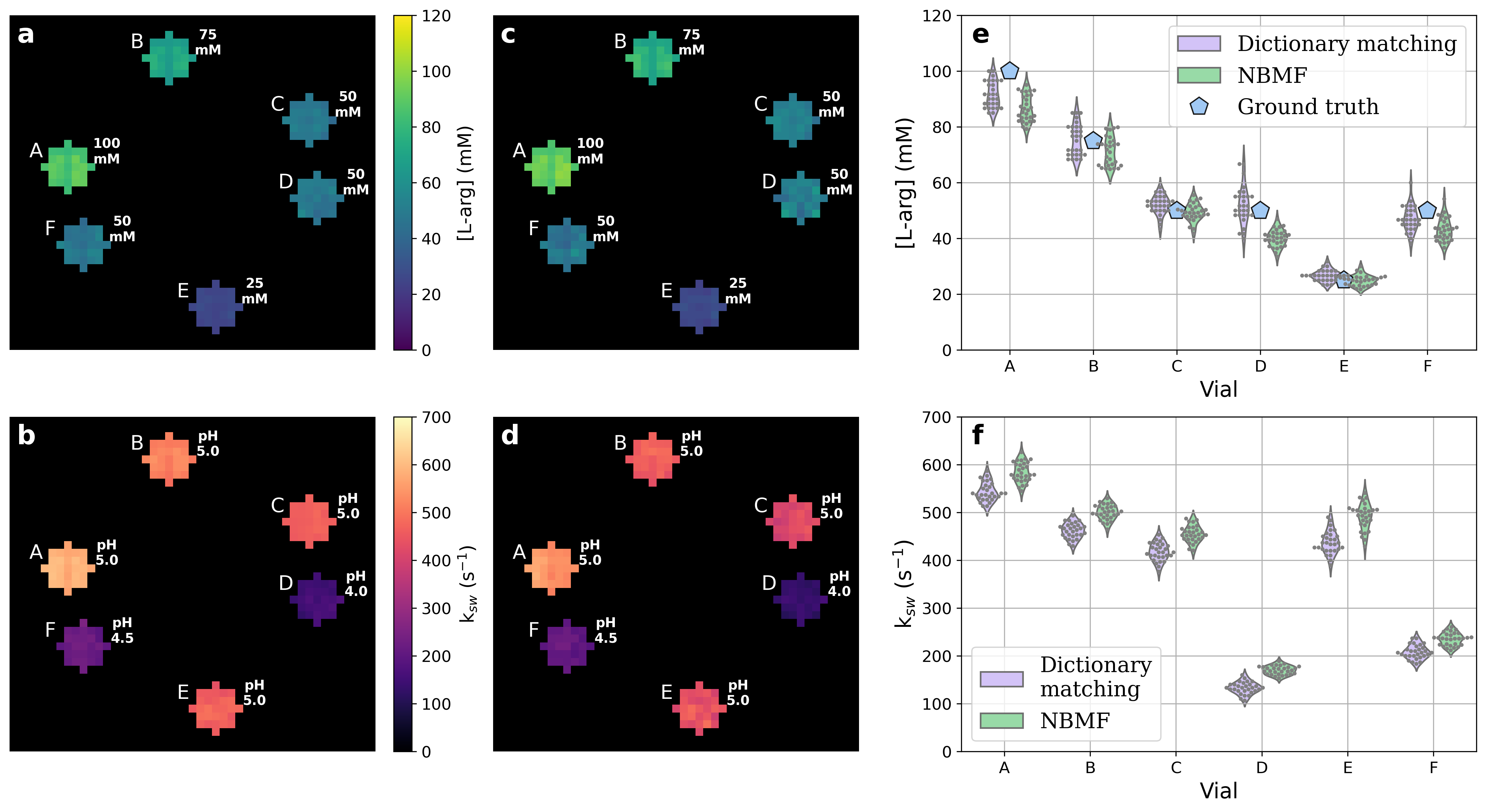}
\caption{In-vitro study. L-arginine samples were imaged using a pulsed CEST-MRF protocol in a 3T clinical scanner. The NBMF-based L-arginine concentrations (\textbf{a}) and proton exchange-rates (\textbf{b}) were in good agreement with those obtained by dictionary-based pattern matching (\textbf{c}, and \textbf{d}, respectively). The ground truth L-arginine concentrations and pH values are mentioned in white text next to each vial. The pixelwise distributions are further compared in (\textbf{e,f}). Each point in the swarm plot reflects a single 1.8 mm x 1.8 mm x 5.4 mm voxel.}
\label{fig:larg}
\end{figure}

A phantom composed of six vials with different combinations of L-arginine concentration and pH was assembled and scanned in a 3T clinical scanner (Prisma, Siemens Healthineers, Germany) using a previously published non-steady-state rapid CEST protocol\cite{Perlman2022_NatureBME_Apoptosis, weigand2023accelerated}. Good agreement was obtained between the NBMF-estimated and known L-arginine concentrations (Fig. \ref{fig:larg}a,e): Pearson’s r = 0.986, p = $3.0 \times 10^{-4}$ (n=6), root mean square error (RMSE) = 8.4mM, mean absolute percentage error (MAPE) = 10.8\%). The NBMF-reconstructed proton exchange rates were in good agreement with the corresponding values estimated by traditional MRF dictionary-matching (Fig. \ref{fig:larg}b,d,f): Pearson’s r = 0.999, p = $1.6 \times 10^{-6}$, RMSE = 41.0 $s^{-1}$, MAPE = 13.2\%. The pH dependence of the NBMF reconstructed exchange rate (Fig. \ref{fig:larg}b) was a good fit for a base-catalyzed proton exchange model (R$^2$ = 0.94, p-value = $1.4 \times 10^{-3}$), as predicted by theory\cite{cohen2018rapid}. An additional in-depth comparison between traditional dictionary matching and NBMF is available in Supplementary Table 1, and Supplementary Figures 1,2.

\begin{figure}[h] {
\centering
\includegraphics[trim={0 0cm 0cm 3.5cm},clip,width=\linewidth]{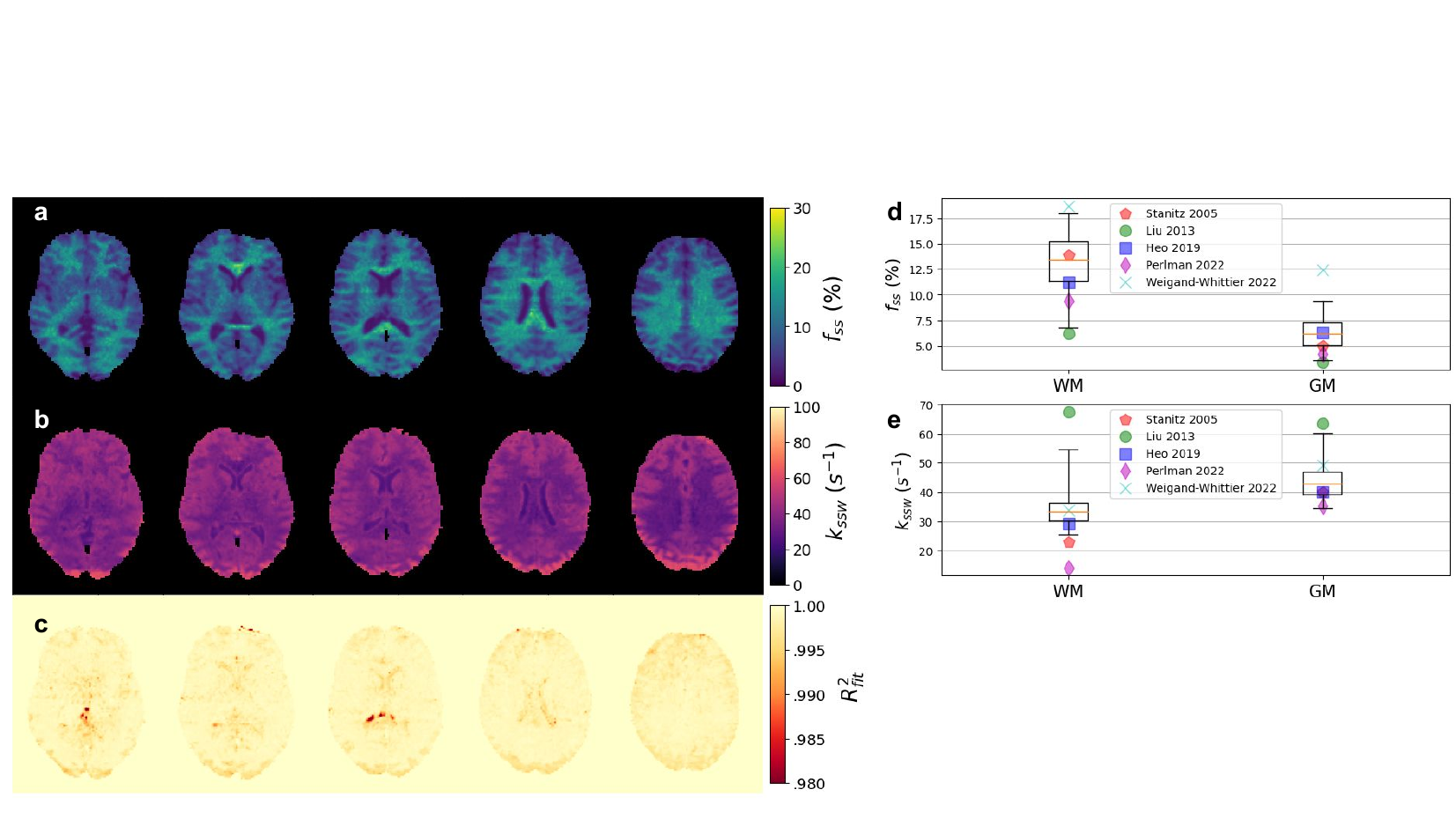}
\caption{NBMF-based quantification of the semisolid MT proton exchange parameters in the healthy in-vivo human brain. \textbf{(a-c)} Representative reconstructed parameter maps of the semisolid-MT proton volume fraction \textbf{(a)} and exchange rate \textbf{(b)}, alongside a fidelity estimation \textbf{(c)} of the data-model agreement, computed as $R^2$=1-NMSE (normalized mean square error).  \textbf{(d-e)} Statistical analysis of the resulting proton exchange parameter values across the brain white matter and gray matter (WM/GM) regions of interest (box-plots, n=47K/65K), compared to literature (colored markers)\cite{Stanisz2005, Liu2013, Heo2019, Perlman2022_NatureBME_Apoptosis, Weigand-Whittier2023}.
} \label{fig:healthy_MT}
} \end{figure}

\subsection*{Quantifying the semisolid-MT and CEST proton exchange parameters in the human brain}
The NBMF pipeline used in-vitro was modified for semisolid-MT and amide proton exchange parameter mapping in the human brain. Two 3D and rapid acquisition pulse sequences were applied, as described previously\cite{Perlman2022_NatureBME_Apoptosis, weigand2023accelerated}. The first sequence varied the saturation pulse frequency offset between 6-14 ppm (designed to encode semisolid-MT information), whereas the second sequence fixed it at 3.5 ppm (for amide proton parameter encoding). In both cases, a total of 31 raw information encoding images were generated, with the saturation pulse power randomly varied between 0-4 $\mu$T\cite{Perlman2022_NatureBME_Apoptosis, cohen2018rapid}. Water relaxation (T$_1$ and T$_2$) and field maps (B$_1$ and B$_0$) were acquired separately and used as an extra input to the neural reconstructor described in Fig. \ref{fig:nbmf_overview} in order to avoid water-pool- and inhomogeneity-associated biases, respectively\cite{Perlman2022_NatureBME_Apoptosis}. A two-step NBMF (see Methods section) was used to fit the semisolid-MT and amide proton exchange parameters to the raw data.

\begin{figure}[h] {
\centering
\includegraphics[trim={0 0cm 0cm 3.5cm},clip,width=\linewidth]{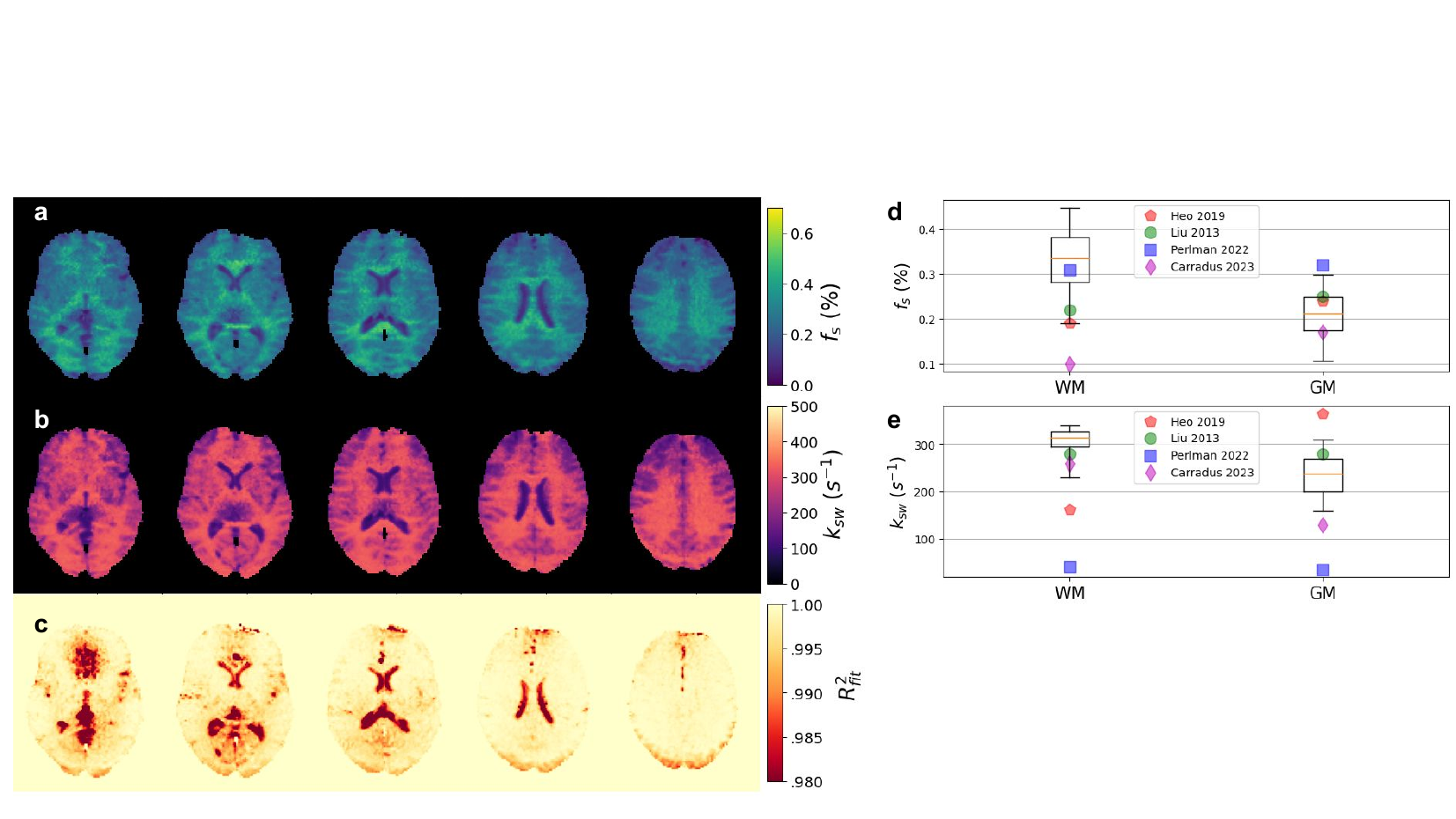}
\caption{NBMF-based quantification of the amide proton exchange parameters in the healthy in-vivo human brain. \textbf{(a-c)} Representative reconstructed parameter maps of the amide proton volume fraction \textbf{(a)} and exchange rate \textbf{(b)}, alongside a fidelity estimation \textbf{(c)} of the data-model agreement, computed as $R^2$=1-NMSE (normalized mean square error).\textbf{(d-e)} Statistical analysis of the resulting proton exchange parameter values, across the brain white matter and gray matter (WM/GM) regions of interest (box-plots, n=47K/65K), compared to literature (colored markers)\cite{Stanisz2005, Liu2013, Heo2019, Perlman2022_NatureBME_Apoptosis, Weigand-Whittier2023}.
} \label{fig:healthy_AMIDE}
} \end{figure}

Quantitative semisolid-MT and amide proton exchange parameter maps derived from a representative healthy volunteer are presented in Fig. \ref{fig:healthy_MT} and Fig. \ref{fig:healthy_AMIDE}, respectively. The resulting proton volume fractions and exchange rates were in agreement with the literature (although the large variability in previous reports is noted; see Fig. \ref{fig:healthy_MT}d-e and Fig. \ref{fig:healthy_AMIDE}d-e). The mean values obtained for white/gray matter (WM/GM) were: $f_{ss}=13.09\pm3.44\ (\%),\ k_{ssw}=34.7\pm7.8\ (s^{-1}),\ f_{s}=0.33\pm0.08\ (\%),\ k_{sw}=305.1\pm34.0\ (s^{-1})$ for white matter and $f_{ss}=6.28\pm1.88\ (\%),\ k_{ssw}=44.2\pm7.5\ (s^{-1}),\ f_{s}=0.21\pm0.06\ (\%),\ k_{sw}=235.9\pm46.0\ (s^{-1})$ for gray matter.

\begin{figure}[h] {
\centering
\includegraphics[trim={0 0cm 2.5cm 0cm},clip,width=\linewidth]{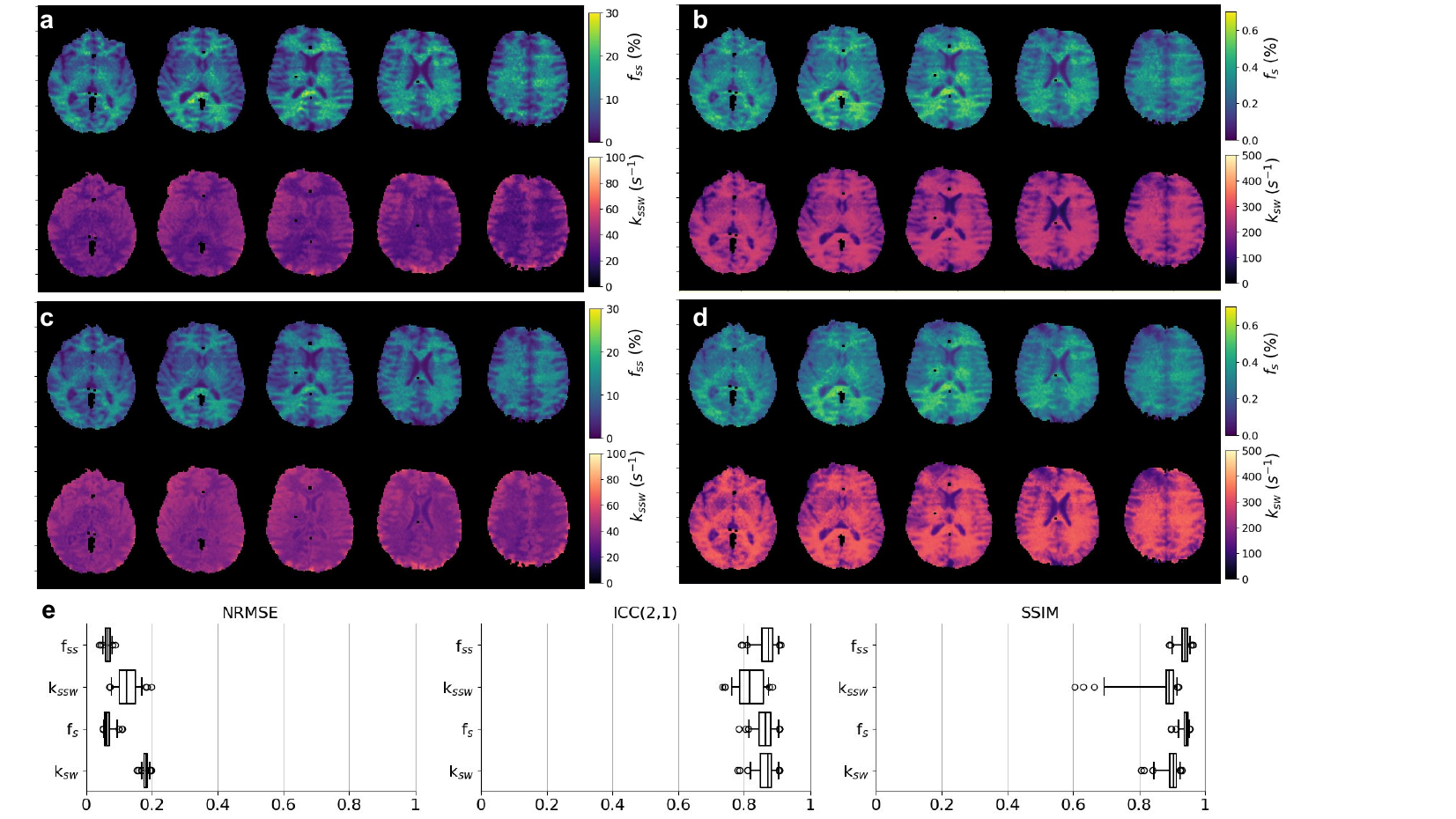}
\caption{NBMF's fitting mode vs. transfer mode. A comparison between the results of single-subject NBMF (\textbf{a, b}) and a real-time quantification of the same subject by inferring the neural reconstructor trained while fitting another one (\textbf{c, d}). A perceptually and quantitatively similar outputs were obtained for both semisolid-MT (\textbf{a}, \textbf{c}) and amide (\textbf{b}, \textbf{d}) parameter mapping. \textbf{e}. 
Similarity analysis using normalized root-mean-square (NRMSE), intraclass correlation coefficient (ICC(2,1), absolute agreement-assessing variant) and structural similarity index measure (SSIM), across all (n=50) processed brain slices.
} \label{fig:X2Y_vs_Y2Y}
} \end{figure}

The joint fit and training of the NBMF produced a neural reconstructor, optimized on a \textit{single} subject. We then re-applied the trained reconstructors to additional subjects in a fast inference mode. A representative example comparing the parameter maps obtained from single-subject NBMF with those obtained by a rapid reconstructor reuse is shown in Fig. \ref{fig:X2Y_vs_Y2Y}. The resulting agreement metrics (Fig. \ref{fig:X2Y_vs_Y2Y}e) were as follows; NRMSE: 7$\pm$1\%, 12$\pm$3\%, 7$\pm$1\%, and 18$\pm$1\%; Intraclass correlation coefficient ICC(2,1): 0.87$\pm$0.03, 0.82$\pm$0.04, 0.86$\pm$0.03, 0.86$\pm$0.03; SSIM: 0.93$\pm$0.02, 0.87$\pm$0.07, 0.94$\pm$0.01, 0.90$\pm$0.03, for the f$_{ss}$, k$_{ssw}$, f$_{s}$, and k$_{sw}$, respectively. Additional analysis is provided in Supplementary Figure 4.

\begin{table}[ht]
\small
    \captionsetup[subtable]{justification=raggedright, singlelinecheck=false} 
    \begin{subtable}[t]{0.9\textwidth}
        \caption{NBMF computational complexity compared to previously reported studies of semisolid-MT and CEST quantification} 
        \begin{tabular}{|p{2.5cm}|p{3cm}|p{3.7cm}|p{3.7cm}|p{2.5cm}|}
            \hline
            \diagbox{Stage}{Method} & Numerical BM model direct fitting & Dictionary-based \newline pattern matching & NN-based mapping \newline(supervised learning) & \textbf{NBMF} (MT+CEST) \\ \hline
            Preparatory steps using synthetic data + 1$^{st}$ subject mapping& None & Hours to days; \newline e.g., 21.3h (x56 nodes) for 79M-entries dictionary generation\cite{Perlman2022_NatureBME_Apoptosis} & Hours; \newline e.g., \textasciitilde4.5h for 60K-entries dictionary generation and network training\cite{cohen2023cest}  & <30min (fit\&train) \\ \hline
            N$^{th}$  subject tissue parameters quantification (N>1) & Hours to days; \newline e.g., 6h recon of a $256\times256$ pixel slice\cite{Heo2019} & Hours to days; \newline e.g., 2.31h for reconstructing a $128 \times 128$ pixel slice\cite{Perlman2022_NatureBME_Apoptosis}  & \textasciitilde1 sec\cite{Perlman2022_NatureBME_Apoptosis,cohen2023cest} \newline (NN inference) 
            & 1sec (inference) \textbf{or}: \newline <30min (re-fitting) \newline  \\ \hline            
        \end{tabular}
        \label{tab:subtable1}
    \end{subtable}
    \vspace{0.2\textwidth}
    \begin{subtable}[t]{0.9\textwidth}
        \caption{Unified benchmarking of all methods using the accelerated approach developed as part of this work\textsuperscript{*}} 
        \centering
        \begin{tabular}{|p{2.5cm}|p{3cm}|p{3.7cm}|p{3.7cm}|p{2.5cm}|}
            \hline
            \diagbox{Stage}{Method} & Autodiff-based (AD) model fitting \textbf{(VBMF)} & Dictionary-based \newline pattern matching & NN-based mapping \newline(supervised learning) & Self-supervised NN + AD \textbf{(NBMF)}  \\ \hline
            Preparation + 1$^{st}$ subject mapping & 3 min & 25 sec$^{\&}$ (generation) +\newline93 sec (matching) & 25 sec (generation) +\newline 58 sec (training) & 3 min (fit\&train) \\ \hline
            $N^{th}$ subject mapping (N>1) & 3 min &93 sec (matching) & 1 sec (NN inference) & 1 sec (inference) 
            \\ \hline \hline
            Consistency with raw per-subject data & \yes \newline(up to convergence) & \yes \newline(up to grid resolution) & \no \newline(empirically, in our test) & \yes \newline(up to prior)
            \\ \hline
            Implicit NN-based smoothing prior & \no & \no & \yes & \yes        
            \\ \hline
        \end{tabular}
        \label{tab:subtable2}
         \raggedright\footnotesize \textsuperscript{*}A GPU-based JAX formulation of the Bloch McConnell ODEs numerical/analytical solutions. The comparison was based on a semisolid MT MRF protocol using a dictionary consisting of 400K entries and a whole-brain quantification task (194K voxels), see additional details in Supplementary Note 2. \\
     \raggedright\footnotesize {$^{\&}$} Dictionary generation for the 79M entries CEST and semisolid MT cartesian grid dictionary used in ref\cite{Perlman2022_NatureBME_Apoptosis} took 80 min, exhibiting a linear scaling of the computational complexity.
    \end{subtable}    
\vspace{-0.2\textwidth}
\caption{Comparative summary of proposed and existing techniques. Computational demands and key optimization features for various methods and implementations of non-steady-state pulsed semisolid-MT/CEST parameter quantification across the entire processing pipeline. The presented times refer to a whole brain reconstruction on a single desktop machine unless noted otherwise. Additional analysis of the compared features is available in Supplementary Notes 1-3.
}
\label{tab:compute}
\end{table}

\subsection*{Computational complexity, timing, and comparison with alternative approaches}
The combined NBMF training and fitting procedure for all four semisolid-MT and amide proton volume fraction and exchange rate parameter maps from the whole brain of a single subject (169K-194K voxels) took $18.3\pm8.3$ minutes on a standard GPU-equipped (GeForce RTX 3060) desktop workstation, of which, the two-pool quantification of the semisolid MT pool parameters took $3.0\pm0.4$ min.  Re-applying the trained reconstructors for whole-brain parameter mapping on a new subject took $1.0\pm0.2$ sec. Overall, the complete quantification pipeline 
takes less than 30 min for NBMF, compared to at least several hours using previously reported implementations of traditional Bloch-Fitting\cite{Heo2019}, or dictionary-based preparation and supervised neural network training.
\cite{Perlman2022_NatureBME_Apoptosis, cohen2023cest} (Table \ref{tab:compute}a).

Next, we performed unified benchmarking of dictionary generation, matching, and supervised learning, using the accelerated approach developed as part of this work (GPU-based JAX formulation of the Bloch McConnell numerical solution); see additional implementation details in Supplementary Note 2.
Notably, this yielded comparable timing to self-supervision (Table 1b), given that a non-cartesian sampling grid is used for dictionary generation. The benefit of nonetheless using the self-supervised approach compared to supervised training is highlighted in Supplementary Notes 1-3 in the context of consistency with the raw acquired data and per-subject discrepancy minimization.  In general, by unlocking rapid direct fitting (via automatic differentiation) and coupling it with self-supervised learning, NBMF constitutes an alternative way to address the ill-posed in-vivo quantification challenge. It contributes to an improved consistency of the quantitative parameter estimates with the raw data given the model, compared to dictionary-based supervised learning (Supplementary Fig. 3). By combining the explicit objective of minimal data-model discrepancy with implicit neural regularization, NBMF created smoother maps with visible contrast, while keeping the data-model agreement close to that achieved by dictionary matching (Supplementary Fig. 5). 

\section*{Discussion}

\subsection*{Automatic differentiation of the Bloch-McConnell (BM) equation solutions.}

Quantification of semisolid MT/CEST proton exchange parameters under non-steady-state conditions is a notable example of a biophysical estimation where the forward model is perceived as too complex for direct inverse solution via fitting (requiring several days for a single whole brain reconstruction\cite{Heo2019}). While the solution can be approximated via MRF, the large simulated signal dictionaries\cite{Cohen2018DRONE, Perlman2022_NatureBME_Apoptosis} associated with multi-pool imaging also demand significant computational resources\cite{Zhu2019,Perlman2022_AUTOCEST}, limiting the development of new pulse sequences. A recently reported dictionary-free alternative\cite{Kang2021}  proposed unsupervised learning for semisolid-MT parameter quantification. However, this method assumes continuous pulse irradiation, which is not available in many clinical scanners, and also relies on analytical solutions, which are not compatible with multi-pool pulsed CEST imaging.

The dictionary-free method presented in this work overcomes all such limitations. Our approach is based on a fundamental insight: by proper formulation, ODE models considered only numerically solvable can become step-wise analytical, and thereby compatible with automatic differentiation-based optimization. Specifically, the suggested formulation enables GPU-based matrix inversion and exponentiation, which translates into efficient gradient descent via back-propagation. Combining this concept with a recently reported high-performing automatic differentiator\cite{jax2018github} provides a new option for solving complex biophysical estimation tasks such as pulsed CEST quantification, demonstrated here. Compared to standard model fitting, this approach avoids (i) computationally expensive and inaccurate purely-numerical derivatives computed via multiple evaluations, and (ii) explicit analytical approximations, which can only be applied to a limited subset of cases and lack generalization (e.g., unsuitable for a multi-proton-pool pulsed-RF saturation). Unlike MRF dictionary-trained networks\cite{Cohen2018DRONE, Perlman2023_Review_w_Heo}, the suggested approach can provide parameter estimates that allow the model to best describe the raw data (Supplementary Fig. 3).

Supervised NN training using a synthetic signal dictionary requires the estimation of the application-specific parameter distribution, which is often unknown in advance. The self-supervised (NBMF) approach circumvents this challenge by training on the in-vivo data itself, offering improved parameter distribution matching. When it comes to re-using the trained network on new unseen subjects, one drawback of this approach lies in its reliance on previously represented proton exchange parameters. Dictionary-based approaches, on the other hand, have the flexibility for representing the expected abnormality values (if they are known) or simply using a very broad parameter distribution that covers both the healthy and diseased states. A future patient cohort investigation is needed to examine the clinical utility of transferring the self-supervised quantification approach, when trained on healthy volunteers, for quantification in unseen pathology (e.g., small lesions).

As shown in Table 1, the most time-consuming steps for supervised dictionary-based learning are the dictionary generation step followed by neural network training. However, if the imaging scenario is a priori known (e.g., brain cancer treatment monitoring) and the acquisition protocol parameters are fixed and optimized, these steps can be done once without affecting the rapid inference time for each new subject. Self-supervised data-based learning (NBMF), on the other hand, offers the flexibility of accommodating various imaging scenarios and is more easily adapted for new acquisition protocols and research directions. That being said, the biophysical modeling developed as part of this work (GPU-based JAX implementation of the Bloch McConnell numerical solution) can also accelerate both dictionary generation and supervised learning (Table 1b), allowing the user to utilize and compare all different approaches. 

The gradient of the forward model can be directly used for simple fitting of the unknown ODE coefficients corresponding to the physical parameters of interest (see voxelwise BMF in the Methods section, Supplementary Note 5, and Supplementary Fig. 7). However, when the core forward model automatic differentiator is also integrated into a self-supervised learning pipeline (NBMF, Fig. \ref{fig:nbmf_overview}), a joint neural representation of the signal-to-parameter transformation can be trained and stored with little extra computational cost. This enables: (i) Faster convergence, which scales well with the number of voxels up to the full brain size, leveraging redundancy towards a spatially smoother solution (Supplementary Figure 7).  (ii) Later reuse for real-time inference on new subjects within a similar imaging scenario. 
 
The human brain imaging results (Figs. 3-5, Supplementary Fig. 3-6,8) reveal the potential for using an autodiff-compatible Bloch-McConnell solver for parameter quantification while training a simple multilayer perceptron (MLP) voxelwise. Combined with a 3D whole brain acquisition routine (which rapidly generates hundreds of thousands of voxels), the suggested system provides a rapid and efficient single-subject learning method. Notably, while this work presented a proof of concept for rapid inference by a network trained on a single subject, robustness and consistency of the transfer leaves a clear room for improvement (Fig. \ref{fig:X2Y_vs_Y2Y} and Supplementary Fig. 4). Future work could study different NN architectures with spatial awareness (via convolutional or attention layers), as well as larger datasets composed of multiple subjects and a combination of both dictionary-based synthetic data and real-world scans. Subsequent efforts could also improve the modeling accuracy by taking under consideration the contributions of additional proton pools (such as amine and guanidinium) to the 3.5 ppm signal.

The proposed approach could be further exploited for other tasks across the CEST-MRF pipeline, such as accelerated dictionary synthesis\cite{vladimirov2024quantitative, nagar2023dynamic}(as demonstrated in Supplementary Note 2 and Table 1b) and pulsed-wave irradiation compatible CEST protocol discovery and optimization\cite{Perlman2022_AUTOCEST, Zhu2019, Loktyushin2021, kang2022learning, Glang2022}. Furthermore, NBMF is directly applicable to anatomical-MRF (proton density, T$_1$, T$_2$) dictionary-free parameter quantification and conversely, to classical non-MRF molecular MRI, such as pulsed multi-B$_1$ Z-spectra fitting\cite{Zaiss2011_LorentzianFit, Zaiss2018_QUESPrev} (see Supplementary Notes 4,5). While auto-differentiation of the Bloch equations has previously been leveraged for several MRI-related applications\cite{zhu2018automated, Perlman2022_AUTOCEST, kang2022learning, lee2019flexible, Luo2021}, to the best of our knowledge this is the first report of utilizing this concept for a \textit{generalized} Bloch-McConnell-\textit{fitting} task. Beyond molecular MRI and MRF, this approach can also be applied to any other diagnostic and biophysical domains that involve dictionary-matching\cite{Ghodasara2020}.

\subsection*{Learning to estimate ordinary differential equation (ODE) coefficients from observed data.}The general strategy underlying NBMF can be applied to any inverse problem that involves fitting ODEs to observations of a dynamical system. In the biomedical realm alone, this 
includes cardiovascular\cite{zenker2007inverse, linial2021_rambamODEcardio}, pharmacokinetic\cite{donnet2013review_SDEpharmakinetics, chou2023_MLpharmakinetics}, and epidemiological\cite{gumel2021_covidODEprimer, keeling2022_covidUKfitting} modeling, among many other tasks.
In parallel to the exponential growth and improvement in AI performance, the last decade has witnessed a surge of interest in harnessing deep learning for physics-based problem solving.
These efforts have most often been directed into two routes: (i) seeking a solution to a partial differential equation (PDE) as an output of a physics-informed neural network (PINN) that operates on spatial and temporal coordinates\cite{Raissi2019, karniadakis2021physicsinformed, Cuomo2022, Rajput2023, Kissas2020}; widely applied for spatially-resolved dynamics in solid\cite{haghighat2021_PINNsolid}- and fluid\cite{cai2021_PINNfluid} mechanics, heat transfer\cite{Cai2021}, power systems\cite{Huang2023}, weather/climate\cite{kashinath2021_PIMLclimate}, and diffusion MRI\cite{Zapf2022}. (ii) Modeling parts of the equation with a NN, yielding a neural ODE/PDE\cite{chen2018_neuralODEs}, often employed as a semi-parametric approach for model discovery\cite{bradley2021_EstimateODEparamsWithNeuralODEs, Lai2021, Rudy2017}. Interestingly, the relatively simple \textit{direct inverse solution} approach described here (Fig. \ref{fig:nbmf_overview}), whereby a NN is trained to infer the coefficients of an ODE model from a few samples of the dynamics, has not received similar attention. This could open opportunities for the current work to inform new approaches to ODE-driven inverse problems across a multitude of tasks.

\subsection*{Conclusions}
The NBMF framework enables rapid, dictionary-free, pulsed-saturation- and multiple proton-pool-compatible semisolid MT/CEST-MRF quantification. By combining a GPU-accelerated auto-differentiable numerical ODE solver and self-supervised deep learning, the NBMF pipeline is able to match alternative AI-reconstruction based inference, while removing the need for dictionary synthesis. 
NBMF is three orders of magnitude faster than traditional Bloch fitting, and provides a one-stop-shop for reconstructing quantitative molecular MRI data. The underlying approach has potential applications across a wide variety of ODE-driven inverse problem tasks.  

\section*{Methods}

\subsection*{CEST Phantoms
}
A set of six 10 ml L-arginine (L-arg, chemical shift = 3 ppm, Sigma-Aldrich) phantoms was prepared at a concentration of 25, 50, or 100 mM. The phantoms were titrated to different pH levels between 4.0-5.0 and placed in a 120 mm diameter cylindrical holder (MultiSample 120E, Gold Standard Phantoms, UK), filled with saline.

\subsection*{Human subjects}
Four healthy volunteers (three females/one male, with average age 23.75$\pm$0.83) were scanned at Tel Aviv University (TAU), using a 3T MRI equipped with a 64-channel coil (Prisma, Siemens Healthineers). The research protocol was approved by the TAU Institutional Ethics Board (study no. 2640007572-2) and the Chaim Sheba Medical Center Ethics Committee (0621-23-SMC). All subjects gave written, informed consent before the study.

\subsection*{MRI Acquisition}
 All acquisition schedules were implemented using the Pulseq prototyping framework\cite{layton2017pulseq} and the open-source Pulseq-CEST sequence standard\cite{Herz2021}. The MRF acquisition protocols were implemented as described previously\cite{Perlman2022_NatureBME_Apoptosis, weigand2023accelerated}, with an unsaturated M$_0$ image added at the beginning of each sequence. A spin lock saturation train (13 $\times$ 100 ms, 50\% duty-cycle) was used for each one of the 30 additional iterations of the sequence, which varied the saturation pulse power between 0 and 4 $\mu$T (average pulse amplitude). The saturation pulse frequency offset was fixed at 3 ppm for L-arginine phantom imaging\cite{cohen2018rapid}, 3.5 ppm for amide brain imaging\cite{Perlman2022_NatureBME_Apoptosis}, or varied between 6 and 14 ppm for semisolid MT imaging\cite{weigand2023accelerated}.
 The saturation block was followed by a 3D centric reordered EPI readout module\cite{mueller2020whole, akbey2019whole}, providing a 1.8 mm isotropic resolution. The in-plane axial matrix size was 116 $\times$ 88, with 50 slices (169K - 194K brain voxels) used per subject. The full sequences can be accurately reproduced using previously published Pulseq (.seq) files\cite{vladimirov2024quantitative}. Each 3D MRF acquisition took 2:36 (min:s). The same readout module was used for acquiring additional B$_0$, B$_1$, T$_1$, and T$_2$ maps, via WASABI\cite{schuenke2017simultaneous}, saturation recovery, and multi-echo sequences, respectively. The total scan time per subject was 9 min. The WASABI sequence used a preparation scheme realized by a rectangular pulse of 5 ms and nominal B$_1$=3.7 $\mu$T. Twenty-four frequency offsets were equally spaced between -1.8 ppm and 1.8 ppm with a recovery time of 4.5 s. An M$_0$ image was taken at -300 ppm with a recovery time of 12 s. The saturation recovery T$_1$ mapping protocol used the following TR (s) values: 10, 6, 4, 3, 2, 1, 0.8, 0.5, 0.4, 0.3,  0.2, 0.1. The T$_2$ mapping multi-echo sequence used the following echo times (s): 0, 0.01, 0.025, 0.03, 0.04 0.05, 0.1, 0.2, 0.3, 0.5, 1.0 with a TR = 10 s.

\subsection*{MRI data pre-processing}
In vitro images with no L-arginine vials, partial vials, or severe image artifacts were removed. Regions of interest (ROIs) were defined using circular masks. In-vivo brain images were motion-corrected and registered using elastix\cite{Klein2010}. WM/GM ROI segmentation of the T$_1$ map was performed using statistical parameter mapping (SPM)\cite{Ashburner2005}. Quantitative reference CEST-MRF values (Fig. \ref{fig:larg}) were obtained using dot-product matching, as extensively described previously\cite{cohen2018rapid, vladimirov2024quantitative}.

\subsection*{NBMF architecture for semisolid-MT and CEST quantification} The self-supervised learning framework comprises two main components (Fig. \ref{fig:nbmf_overview}, Top): 

(A) Reconstructor $\mathcal{R}$ - a fully-connected multi-layer perceptron (MLP) NN, applied voxel-wise on the raw input data\cite{Cohen2018DRONE, Perlman2022_NatureBME_Apoptosis, Cohen2023, power2024vivo}. The NN is composed of three layers, with 256 neurons and ReLU activation in each hidden layer. The output layer consists of 5 neurons, encoding the estimates for the proton volume fraction and exchange rate of the compound of interest and their joint uncertainty expressed as noise covariance. It utilizes a sigmoid activation, with the output scaled to a predefined range of the parameter values, which effectively defines the prediction boundaries as follows: semisolid proton volume fraction f$_{ss}\in$[0, 30] (\%), semisolid proton exchange rate k$_{ssw}\in[0, 150]$ (s$^{-1}$), amide proton volume fraction f$_{s}\in$[0, 1.2] (\%), amide proton exchange rate k$_{sw}\in$[0, 400] (s$^{-1}$), L-arginine concentration [L-arg]$\in[10, 120]$ (mM) and N$_{\alpha}$-amine (of L-arginine) proton exchange rate k$_{sw}\in$[100, 1400] (s$^{-1}$)\cite{cohen2018rapid}. Several auxiliary maps \textbf{X}, including water relaxation T$_1$, T$_2$, and B$_0$/B$_1$ inhomogeneities, are appended to the MRF raw data \textbf{D} to be used as inputs for the tissue parameter estimation: $\Tilde{\textbf{P}}=\mathcal{R}\left((\textbf{D}, \textbf{X}), w\right)$ where $w$ are the weights to be trained.

(B) Simulator $\mathcal{F}$ - a differentiable multi-pool spin physics solver. A numerical simulation of the piecewise-constant coefficient Bloch-McConnell (BM) differential equations was implemented in the open-source JAX\cite{jax2018github} computing framework, leveraging its strong auto-differentiation and GPU-acceleration capabilities for matrix operations.
The simulator concatenated and chained the calculations of the BM closed-form solution across all pulses and delays of the protocol. This was carried out by inversion and exponentiation of the 9 $\times$ 9 BM-matrix $\mathbf{A}$, which expresses all precession, saturation, relaxation and exchange terms of the multi-pool magnetization vector ($\mathbf{M}$) dynamics, as previously defined\cite{Roeloffs2015}: 
\begin{equation}
\label{eq:evolution} 
\hspace{50pt}
    \mathbf{\dot M} = -\mathbf{AM + C} \hspace{3pt}\xrightarrow{}\hspace{7pt} \mathbf{M}_{eq.} = \mathbf{A\backslash C} \hspace{3pt};\hspace{7pt} \mathbf{M}(t+\Delta t) = e^{-\mathbf{A} \Delta t} \left(\mathbf{M}(t)-\mathbf{M}_{eq.}\right) + \mathbf{M}_{eq.}    
\end{equation}
This solver is compatible with the rectangular pulse-train shape employed in this study and others\cite{Perlman2022_NatureBME_Apoptosis, weigand2023accelerated, Herz2021}, while arbitrary pulse shapes can be supported using a simple matched-RMS approximation, or to any order
through a Magnus expansion\cite{BLANES2009}. For two-pool imaging cases (such as semisolid MT data acquired using saturation pulses with a frequency offset higher than 6 ppm), additional acceleration was obtained by implementing the interleaved saturation–relaxation (ISAR2) approximate analytical solution\cite{Roeloffs2015} of the saturation stage. The RF pulses of spin-lock and readout flips were approximated as hard pulses generating precise flip angle rotations. 

The model is designed to represent the whole sequence by simulating the Z-magnetization dynamics during the recovery, saturation, and readout stages, provided that spoilers are applied. For each of the two (semisolid MT/amide) sequences, the Nx31 non-steady-state MRF measurements from 169K-194K brain voxels were normalized using an unsaturated M$_0$ reference image. Thus, the resulting acquired data $\textbf{D}=\{D_n\}_{n=1}^N \in[0, 1]$ is directly related to the magnetization vector governed by Eq. \eqref{eq:evolution}, at the end of the saturation pulse block. Therefore,
given $\Tilde{\textbf{P}}$, an estimate of the sought parameters, the simulator provides a re-synthesis of the data as: 
$\Tilde{\textbf{D}}=\mathcal{F}(\Tilde{\textbf{P}},\textbf{X},\omega_{rf},B_1$)
, where $\omega_{rf}$ and B$_1$ are the saturation pulse frequency offsets and powers implemented in the MRF protocol, and \textbf{X} are any known tissue parameters.

The NBMF reconstruction of the semisolid-MT proton exchange parameters from the first (\textbf{1}) subject, was obtained by using the MT-MRF data $\mathbf{D}_{ss}^{(1)}$ alongside independently quantified auxiliary parameter maps $\mathbf{X}_{w,B}^{(1)}=\{T_{1w},T_{2w},B_1,B_0\}$,
for training the weights $w_{ss}^{(1)}$ of a neural reconstructor $\mathcal{R}_{MT}^{(1)}$, designed to quantify the associated proton exchange parameters ($\mathbf{\Tilde{P}}_{ss}^{(1)}={f_{ss}, k_{ssw}}$). 
To that end, the NBMF optimizes the following self-supervised objective of consistency with the biophysical model $\mathcal{F}$:
\begin{equation}    
\hspace{50pt} w_{ss}^{(1)} = \underset{w}{\argmin} \left|\left|\Tilde{\textbf{D}}-\mathbf{D}_{ss}^{(1)} \right|\right| = 
\underset{w}{\argmin} \left|\left|\mathcal{F}\left(\mathcal{R}\left((\mathbf{D}_{ss}^{(1)},\mathbf{X}_{w,B}^{(1)}), w\right),\mathbf{X}_{w,B}^{(1)}\right)-\mathbf{D}_{ss}^{(1)} \right|\right| 
\label{eq:NBMF}
\end{equation}
The $L_2$ norm was used as the regression loss. A cosine-decay learning rate schedule and simple early-stopping upon convergence (loss trend reaching plateau) were applied. Augmentation by noise was applied, twice: (a) Adding a $\pm$0.1\% Gaussian noise to the raw samples (b) Adding a Gaussian noise to the $\Tilde{f},\Tilde{k}$ tissue parameters estimate, using covariance derived from extra outputs of the NN, inspired by a recent work\cite{Glang2020}.

This process was repeated using the non-steady-state amide raw MRF data $\mathbf{D}_{s}^{(1)}$ for NBMF quantification of the amide proton exchange parameters ($\mathbf{P}_s={f_{s}, k_{sw}}$). For human brain experiments, we also appended the semisolid MT pool parameter estimates $\Tilde{f}_{ss}$ and $\Tilde{k}_{ssw}$ (obtained from the semisolid MT NBMF procedure) to the auxiliary vector $\mathbf{X}$. This vector served as input for the amide reconstructor $\mathcal{R}_{s}$ and the 3-pool biophysical model $\mathcal{F}_{s}$, so that: $ X_{B,w,ss} = \{X_{B,w}, P_{ss}\} = \{T_{1w},T_{2w},B_1,B_0, \Tilde{f}_{ss}, \Tilde{k}_{ssw}\}$\cite{Perlman2022_NatureBME_Apoptosis}. For the two-pool L-arginine phantom experiments, the auxiliary parameters were assigned constant values based on previous reports (T$_{1w}$ = 2800 ms, T$_{2w}$=1200 ms)\cite{weigand2023accelerated}. 

Importantly, we obtain both the subject-specific proton exchange parameters $\mathbf{\Tilde{P}}^{(1)}=\mathcal{R}^{(1)}(\mathbf{D}^{(1)},\mathbf{X}^{(1)})$ and the trained reconstructor $\mathcal{R}$ at the convergence of the NBMF. This enables ultra-fast quantification of the proton exchange parameters $\mathbf{\Tilde{P}}^{(2)}=\mathcal{R}^{(1)}(\mathbf{D}^{(2)},\mathbf{X}^{(2)})$ from a new subject (\textbf{2}) (Fig. \ref{fig:nbmf_overview} bottom). Notably, this rapid inference is only applicable for new data drawn from the same distribution and cannot be applied to entirely new systems (such as muscle creatine quantification using brain-data trained NBMF). 

As a natural ablation of the system by removing the neural component, the auto-diff simulator can be used for direct voxelwise parameter fitting:
$\mathbf{\Tilde{P}}^{(1)} = argmin ||\mathcal{F}\left(\mathbf{P}^{(1)}\right)-\mathbf{D}^{(1)} ||$, referred to here as voxelwise Bloch-McConnell fitting (VBMF). This simpler process can be described in the context of Fig. \ref{fig:nbmf_overview} as stopping the gradients at the tissue parameters, which now assume the role of independent per-voxel variables. Apart from the obvious drawback of not yielding a neural reconstructor, VBMF's performance is inferior to NBMF for brain imaging (Supplementary Fig. 6), which we ascribe to the implicit smoothing regularization by the neural network.
However, it is a viable direct method for in vitro analysis that is equally able to converge to the minimum of the modeling-error landscape (Supplementary Figs. 1,2).

Finally, additional acceleration was achieved by parallelization of the computational graph across consecutive readout pairs $\{{D_{n-1}, D_{n}}\}$, decoupling the single-iteration simulators $\mathcal{F}_n$. 
Assuming that the $D_{n-1}$ snapshot captures the preceding spin history evolution, the re-synthesis stage is now formulated as $\Tilde{\textbf{D}}=\{\mathcal{F}_n(\Tilde{\mathbf{P}}, D_{n-1})\}_{n=1}^N$ and embedded in Eq. \eqref{eq:NBMF} as such. See Supplementary Note 4 for further elaboration.

 \subsection*{Statistical analysis}
The SSIM and ICC(2,1) were calculated using the open-source SciPy and Pingouin scientific computing libraries for Python. In slice-statistic box plots (Fig. \ref{fig:X2Y_vs_Y2Y}e), the central horizontal lines represent median values, box size represents the two central (2nd, 3rd) quartiles, whiskers represent 1.5 $\times$ the interquartile range above and below the upper and lower quartiles, and circles represent outliers. In the voxel-statistic box plots (Figs. \ref{fig:healthy_MT}, \ref{fig:healthy_AMIDE}) the whiskers represent the 90 central percentiles and outliers are omitted. Numerical results in the text are presented as mean$\pm$SD.

\subsection*{Data availability}
The data for the phantom experiment and sample human brain datasets (a single slice for each of the n=4 subjects) are available at \url{https://github.com/momentum-laboratory/neural-fitting} and \url{https://zenodo.org/records/15021550}. The complete 3D human data cannot be shared due to subject confidentiality and privacy.

\subsection*{Code availability}
The code used in this work is available at \url{https://github.com/momentum-laboratory/neural-fitting} and \url{https://zenodo.org/records/15021550}.


\section*{Acknowledgements}
The authors thank Tony Stöcker and Rüdiger Stirnberg for their help with the 3D EPI readout. The study was supported in part by MOONSHOT-MED, a joint grant program from Tel Aviv University and Clalit Innovation, the innovation arm of Clalit Health Services, and the Edmond J. Safra Center for Bioinformatics at Tel Aviv University. This work was supported by the Ministry of Innovation, Science and Technology, Israel, and a grant from the Blavatnik Artificial Intelligence and Data Science Fund, Tel Aviv University Center for AI and Data Science (TAD). This project was funded by the European Union  (ERC, BabyMagnet, project no. 101115639). Views and opinions expressed are, however, those of the authors only and do not necessarily reflect those of the European Union or the European Research Council. Neither the European Union nor the granting authority can be held responsible for them. 

\section*{Author contributions statement}
A.F. and O.P. conceived the computational framework with N.V. providing valuable input. M.Z., N.V. and O.P. developed and/or conducted the imaging protocols and data acquisition. A.F. was responsible for the AI design, optimization, and data analysis. A.F. wrote the manuscript. O.P. supervised the project. All authors reviewed the manuscript. 

\setcounter{figure}{0}
\setcounter{table}{0}
\captionsetup[figure]{name=Supplementary Figure}
\captionsetup[table]{name=Supplementary Table}
\pagebreak
\section*{Supplementary Note 1.  Bias and uncertainty analysis in vitro}
The data obtained from the controlled phantom experiment provides a convenient opportunity for a deeper comparative analysis between different CEST quantification methods due to the (A) partial availability of ground truth values, (B) good agreement between the well-controlled data and model, and (C) lightweight computational load. We start by addressing the general question of the quantification uncertainty at any given voxel via a deeper use of the signals dictionary. The "dot-product" pattern matching, as typically used in MRF, provides unused information about the actual error values obtained - for the best match and all other points of the parameter-space grid. Assuming $l_2$-normalized signals and given a quantification parameter candidate pair $\{f,k\}$ (proton volume fraction and exchange rate), the experimentally \textit{measured} signal $\textbf{m}$ is modeled by the \textit{simulated} dictionary entry $\textbf{s}$, as $\textbf{s}(f,k)=\textbf{m}+\textbf{e}$. The dot-product (the maximal value of which is sought in the standard MRF procedure) can be related to the error \textbf{e} using: $1-\left<\textbf{m}, \textbf{s}\right>\approx\frac{1}{2}||\textbf{e}||^2 $. The latter value is the Normalized Root-Mean Square Error (NRMSE), where the normalization is by the signal's $l_2$ norm. It can be argued that small differences in the fitting error between the best-match and \textit{other} parameter-pair entries can be interpreted as high uncertainty. Thus, for each voxel, a parameter-space map $||\textbf{e}(f, k)||$ of this "modeling NRMSE" can provide a measure of uncertainty and reflect the bias generated by the physical model itself (if the ground truth is known, as in the in vitro case). In addition, quantification by any non-dictionary method can be projected on top the parameter uncertainty space as a point prediction. In Supplementary Fig. \ref{SIfig:uncertainty} we provide such an analysis for the in-vitro experiment by analyzing a central voxel from each L-arginine vial. For creating the parameter uncertainty space, traditional synthetic signal dictionaries were created using the open-source code provided in ref.\cite{vladimirov2024quantitative}: https://github.com/momentum-laboratory/molecular-mrf. 

\begin{figure}[b!]
\centering
\includegraphics[trim={0 0 0 0cm},clip,width=0.95\linewidth]{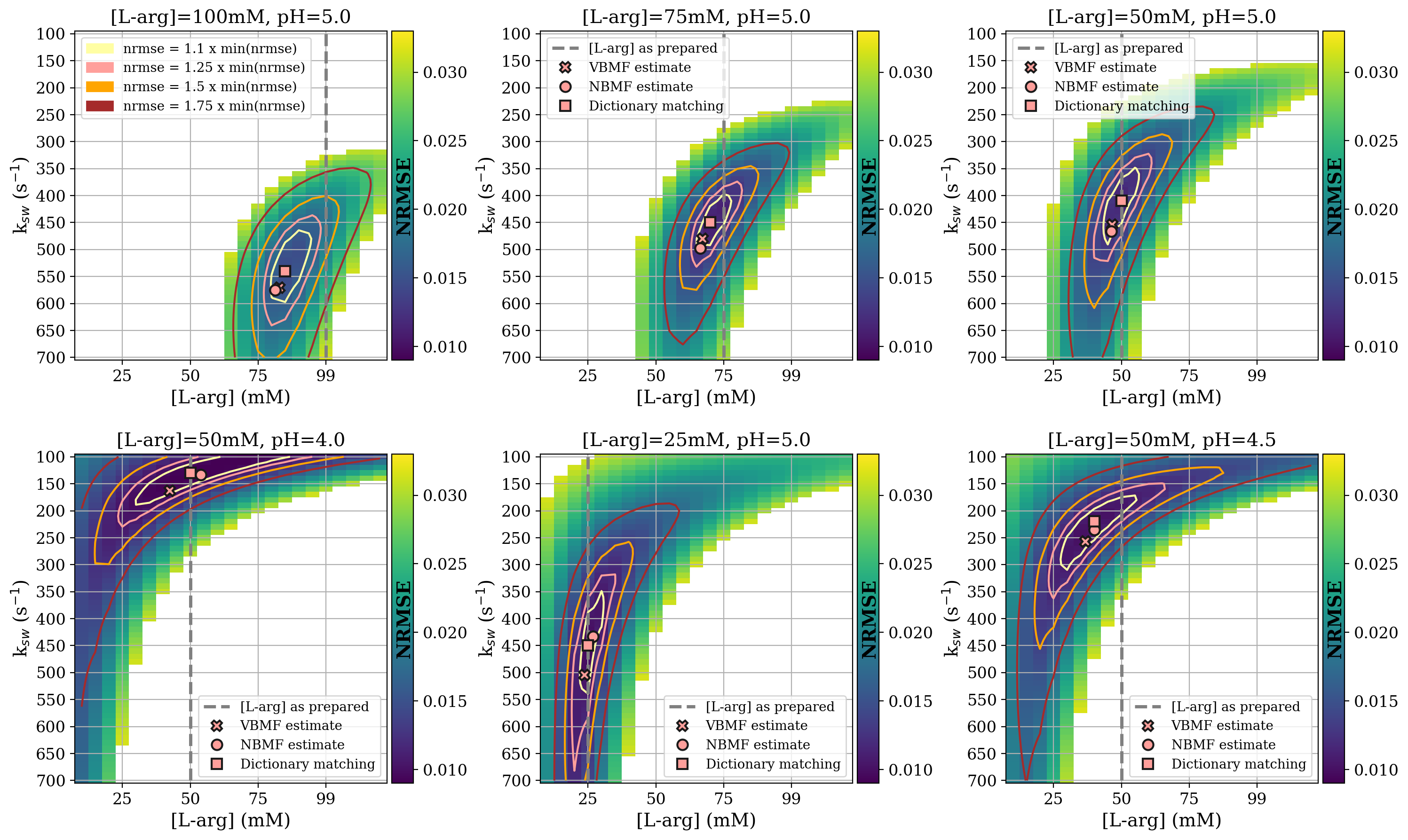}
\caption{\textbf{Single-voxel uncertainty and bias analysis in-vitro}. Each panel describes a single voxel taken from the center of the respective L-arginine vial shown in Fig. 2. Dot-product matching was performed with all dictionary entries in the \{[L-arg], k$_{sw}$\} parameter space to create modeling error maps, dictionary-based minima estimates, and NRMSE at c$\times$NRMSE$_{min}$, c=1.1, 1.25, 1.5, 1.75.
} \label{SIfig:uncertainty}
\end{figure}

As shown in Supplementary Fig. \ref{SIfig:uncertainty}, quantitative estimates of L-arginine concentration and amine proton exchange rate (k$_{sw}$), by either NBMF and VBMF generally fall within the NRMSE < 1.1 $\times$ NRMSE$_{min}$ "confidence region", and are mostly similar between the two variants. Crucially, this analysis serves to highlight the challenge presented by the ill-posed, nonlinear, multi-dimensional quantification problem. The dependence of the simulated signal and subsequently the modeling error on some of the parameters is, on some cases, insufficiently strong (as visualized by the relatively wide confidence regions), destabilizing min-MSE estimation. This effect is quite notable even in the controlled in-vitro case, which presents a good agreement between the data and the model: a NRMSE of best estimates in the 0.01-0.02 range (1-2\%). In the in-vivo case, additional pools and possibly other unaccounted-for physical phenomena lead to a modeling NRMSE which is often above 3\%, as shown in the next section. 

Additional ways for comparing the quantitative parameter estimates of NBMF, VBMF, and dictionary matching (now across all image slice voxel) are presented in Supplementary Fig. \ref{SIfig:phantom_BA} and Table \ref{tab:in_vitro_mean_std}. As shown in Supplementary Fig. \ref{SIfig:phantom_BA}, both NBMF and VBMF results are in good agreement with the dictionary matching output  (Pearson's r $>$ 0.983, p $<$ 10$^{-6}$). While the estimates sometimes differ, not much systemic bias is observed beyond what could be reasonably expected from the modeling-error landscape for the problem as analyzed for sample voxels in Supplementary Fig. \ref{SIfig:uncertainty}. Note how the discrepancies in the two quantitative parameters estimated are anti-correlated (when the L-arginine concentration is underestimated compared to the ideal y=x line, the proton exchange rate is overestimated, and vice versa). This effect is generally in line with the eccentric rotated uncertainty regions as visualized in Fig. \ref{SIfig:uncertainty}, reflecting the basic physics of the quantification task - the CEST effect is roughly proportional to f$_{s} \times $k$_{sw}$. 
\begin{figure}[ht]
\centering
\includegraphics[trim={0 0 0 0cm},clip,width=0.95\linewidth]{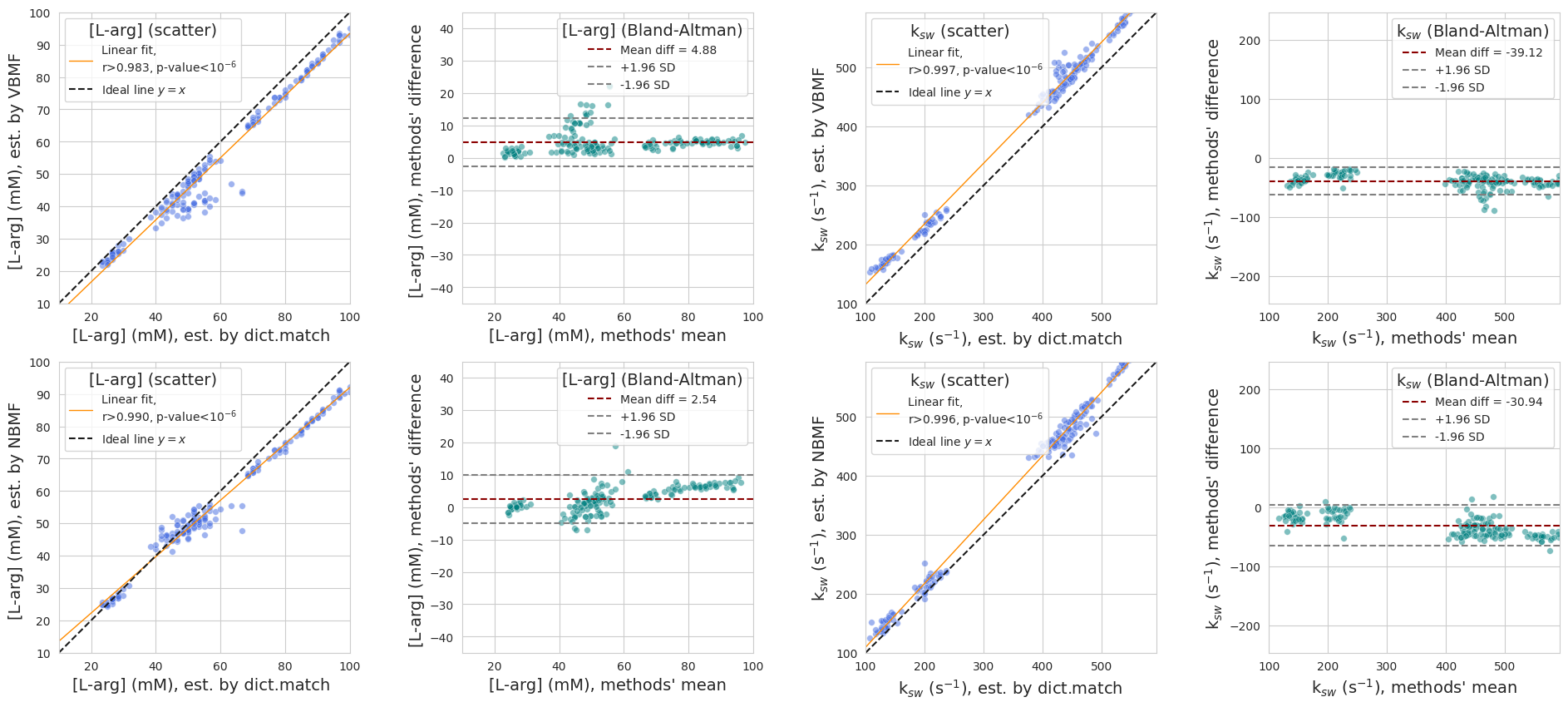}
\caption{\textbf{In-vitro CEST experiment analysis}. Comparing VBMF (\textbf{top}) and NBMF (\textbf{bottom}) quantitative estimates of concentration (\textbf{left}) and exchange rate (\textbf{right}) to those obtained using traditional dictionary matching\cite{vladimirov2024quantitative}. 
} \label{SIfig:phantom_BA}
\end{figure}

\begin{table}[ht]
\begin{tabular}{llllll}
\toprule
\multicolumn{1}{r}{Parameter$\rightarrow$} & 
\multicolumn{2}{c}{[L-arg] (mM)} & 
\multicolumn{2}{c}{k$_{sw}$ (s$^{-1}$)} \\
\multicolumn{1}{l}{Vial$\downarrow$} & 
\multicolumn{1}{r}{Estimator$\rightarrow$} & 
Dictionary & NBMF & Dictionary & NBMF \\
\midrule
A (100mM, pH=5.0) & 91.7 ± 4.6 & 86.6 ± 4.7 & 543.2 ± 19.5 & 584.1 ± 21.0 \\
B (75mM, pH=5.0) & 75.5 ± 5.7 & 71.2 ± 5.0 & 462.0 ± 16.0 & 499.8 ± 16.0 \\
C (50mM, pH=5.0) & 52.1 ± 3.4 & 49.4 ± 3.2 & 417.9 ± 18.7 & 455.0 ± 17.1 \\
D (50mM, pH=4.0) & 52.0 ± 6.6 & 40.1 ± 3.1 & 132.9 ± 12.2 & 170.2 ± 9.1 \\
E (25mM, pH=5.0) & 26.7 ± 2.0 & 25.0 ± 1.9 & 437.6 ± 23.5 & 491.3 ± 27.6 \\
F (50mM, pH=4.5) & 47.0 ± 4.5 & 43.2 ± 4.1 & 208.2 ± 14.3 & 235.3 ± 13.5 \\
\bottomrule
\end{tabular}
\caption{\textbf{Comparing in-vitro parameter estimates by NBMF to those obtained using dictionary-matching.}}
\label{tab:in_vitro_mean_std}
\end{table}

\pagebreak
\section*{Supplementary Note 2. A comparison between self-supervised fitting (NBMF) and dictionary-based supervised  learning in vivo}
In traditional biophysical modeling, parameter estimates are commonly obtained by fitting the experimental data to a physical signal model and minimizing the difference between the measured and the model-generated signal. That difference constitutes the fitting error, which reflects the goodness of fit and the uncertainty region. This work re-introduces fitting into CEST-MRF, a less common approach due to the associated computational load (which is addressed here using auto-differentiation). The error metrics evolve as we transition to machine learning approaches. Dictionary-based learning\cite{Cohen2018DRONE, Perlman2022_NatureBME_Apoptosis}, a more common reconstruction route in MRF, typically focuses on the error of the quantitative estimates themselves. This error is explicitly minimized in the supervised training and can be gauged on a held-out (yet also synthetic) dataset. When the pre-trained neural network is used to infer the biophysical parameters from an experimental signal directly, no fitting error is computed. However, validating the predictions remains crucial. In the absence of ground truth (the in-vivo case), one can still assess the results in the context of the physical model assumed for training. Thus, the natural approach to evaluate the accuracy of the (neural network-based) prediction is to process its output through the forward model of the physics and compare it to the actual measured signal. The discrepancy observed in this step can be described as the \textbf{modeling error}, serving as a crucial validation metric. 
  
\begin{figure}[bh!]
\centering
\includegraphics[trim={0 0 0 0.2cm},clip,width=0.95\linewidth]{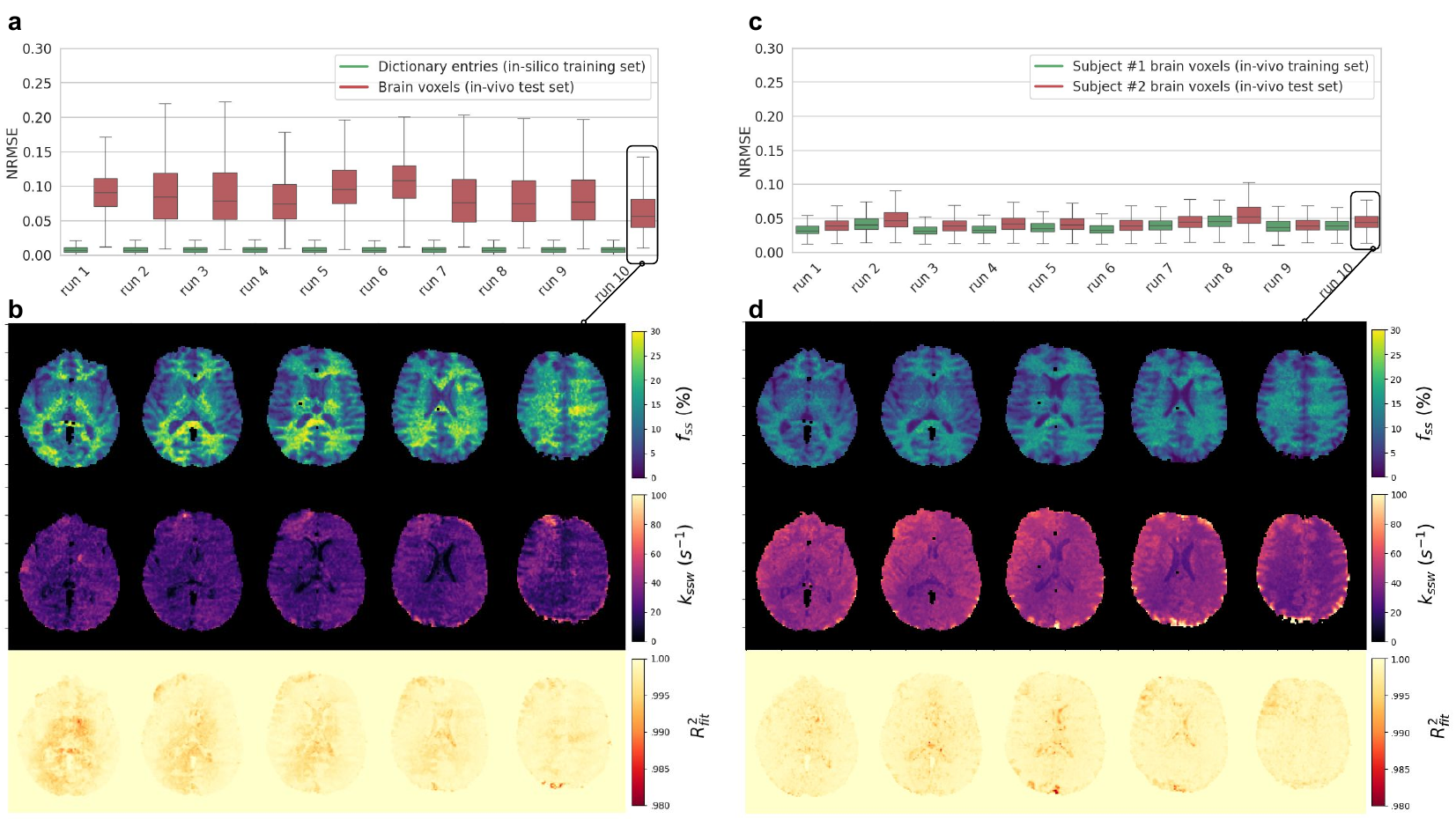}
\caption{\textbf{Robustness analysis}. \textbf{(a, c)}: A comparison between the voxelwise training/test modeling errors of dictionary-based supervised learning (\textbf{a}) and self-supervised learning (NBMF) (\textbf{c}). The modeling NRMSE capturing the circuit error of processing the normalized measured signal $\mathbf{m}$ through the neural reconstructor $\mathcal{R}$ and the physical forward model $\mathcal{F}$ (applied on the resulting parameter estimates $\mathcal{R}(\textbf{m})$) is computed as $||\mathcal{F}\left(\mathcal{R}\left(\mathbf{m}\right)\right)-\mathbf{m}||$, aggregated across training/test set entries and shown for 10 runs of the pipeline (differing in randomness realizations at initialization and training noise injection). \textbf{(b, d)}: Quantitative parameter estimates $\{f,k\}=\mathcal{R}(\textbf{m})$ for five representative slices alongside R$_{fit}^2$=1-NRMSE$^2$ maps. For the supervised learning (\textbf{b}), the reconstructor with the best post-hoc test results out of $\{\mathcal{R}_j;j=1..10\}$ is selected, while for NBMF (\textbf{d}), a typical run is used.
} \label{SIfig:training_robustness}
\end{figure}

In the previous section, we performed a single-voxel analysis of such a modeling error, specifically a \textbf{modeling-NRMSE} for estimates obtained by NBMF/VBMF alongside any possible values in the restricted parameter space of a two-pool problem with T$_1$, T$_2$ fixed. This level of detail and analysis is less feasible for the multi-parameter and more complex in-vivo case. However, it is possible and, in fact, imperative to analyze the modeling error that emerges from the application of a neural reconstructor trained to predict the quantitative parameters from the MRF signal. Keeping with the format of Figures 3 and 4 from the main text, we present (Supplementary Fig. \ref{SIfig:training_robustness}) a voxelwise measure of the goodness of fit as an R$^2_{fit}$ analog, defined as 1-NRMSE$^2$ (based on the differences between the raw experimental image data and the quantified parameters fed into the forward model to re-generate the raw data). 

For the comparative benchmark, we implemented dictionary generation using the same JAX-based GPU-accelerated simulation engine as used in NBMF, following the existing best practices as described in recent works\cite{cohen2023cest, Perlman2022_NatureBME_Apoptosis, Perlman2023_Review_w_Heo, vladimirov2024quantitative}. A three-pool dictionary of 400,000 entries was generated, with varied Latin Hypercube sampled T$_1$, T$_2$, B$_0$, B$_1$, k$_{ssw}$, and f$_{ss}$ in the parameter range described in previous works \cite{Perlman2022_NatureBME_Apoptosis, cohen2023cest}. The generation took 25 seconds. The dictionary-based supervised learning was then performed using a fully connected neural network with the same architecture and predictive power as the NBMF. Network training took 58 seconds per 100 epochs, which was suitable for convergence. The simulated signals used as inputs for the supervised learning were injected with added white Gaussian noise of 2.5\%, empirically found to yield the best data consistency results. Ten repeated training runs were performed, each testing the resulting neural reconstructor on the in-vivo human brain data obtained from healthy volunteer \#1. A similar repeated-run process was applied for NBMF, performing self-supervised training on data from volunteer \#2 and using the obtained neural reconstructor on volunteer \#1.  The quantitative parameter maps obtained by each reconstructor were processed through the forward physics simulator to calculate the modeling NRMSE per voxel as described above.

The results for each of the 10 runs are presented in Supplementary Fig. \ref{SIfig:training_robustness} \textbf{(a, c)}, alongside five representative slices \textbf{(b,d)} of a single reconstruction by the NN trained in one of the runs, chosen as the one with the best test-set modeling-NRMSE achieved by the dictionary-based supervised learning approach (marked using a black rectangle in \textbf{(a)}). 
For supervised learning, it is evident that while the voxelwise modeling error appears robust across repeated runs when evaluated on the synthetic dictionary-based training set (green box plots), it increases dramatically and somewhat unpredictably when the neural reconstructor is applied to real-world subject data (red box plots). On the other hand, self-supervised learning (NBMF) achieves lower modeling errors that are roughly similar between training (green box plots) and test (red box plots) subjects. While the robustness is imperfect, the typical run yields smooth quantitative parameter maps with WM/GM values in line with the literature (see Figure 3 in the main text). Notably, the modeling error both before and after transfer to a different subject is not far from the lowest possible as obtained by dictionary matching (see next section). Thus, NBMF produces voxelwise parameter estimates consistent with the raw data (even after transfer to a different subject, albeit to a lower degree), while reflecting an implicit smoothing regularization via the neural-mapping constraint (see next section). Additional quantitative analysis of self-supervised NN transfer ability by probing agreement between transferred and subject-specific retrained reconstructor is provided in Fig. 5 in the main text (slice-wise) and in Supplementary Fig. \ref{SIfig:braintransfer_BA} (voxel-wise).

\begin{figure}[bt]
\centering
\includegraphics[trim={0 0 0 0cm},clip,width=0.95\linewidth]{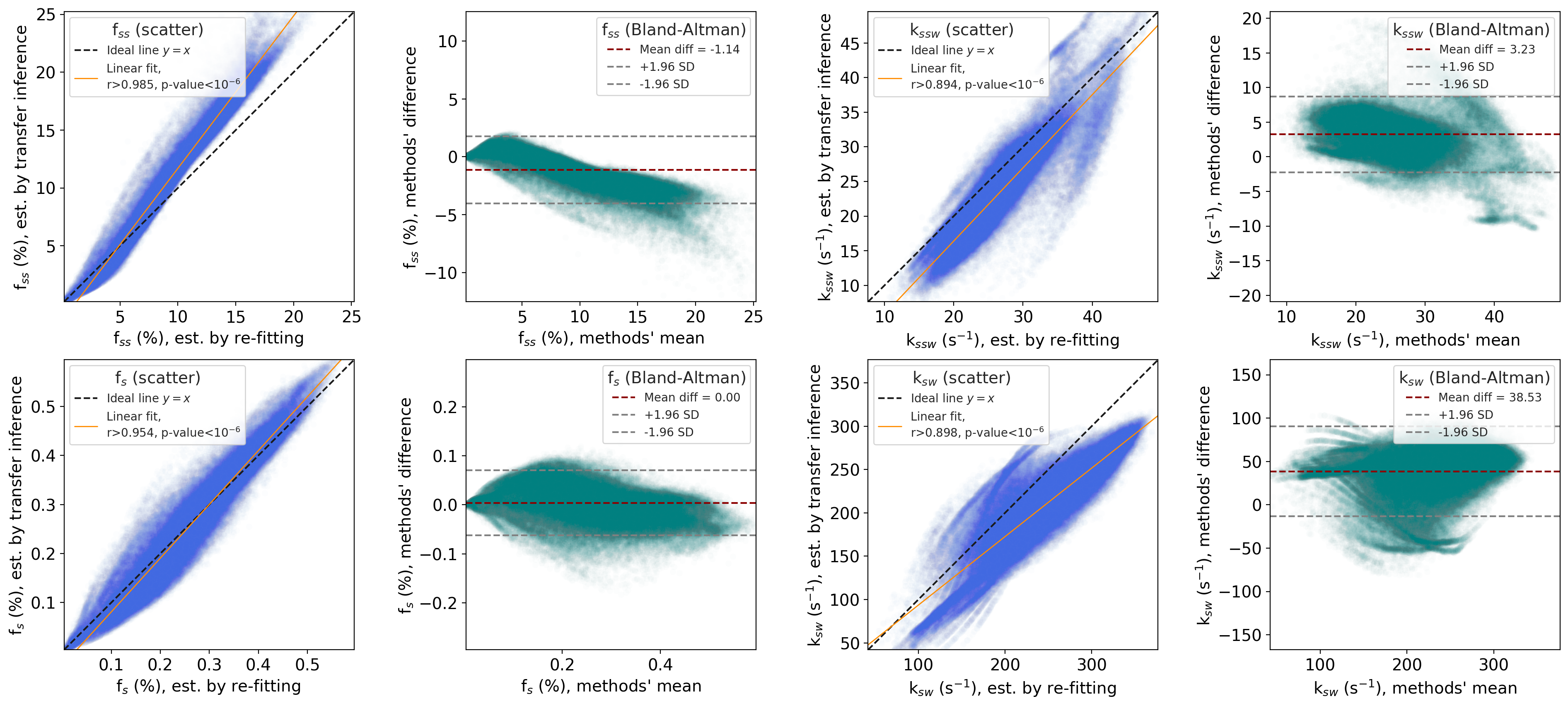}
\caption{\textbf{Fitting vs. Transfer}. Comparing the two modes of NBMF operation (controlled fitting and neural inference) for in-vivo human brain semisolid-MT (\textbf{top}) and CEST-MRF (\textbf{bottom}) quantification. Estimated proton volume fraction (\textbf{left}) and exchange rate (\textbf{right}) values are shown. 
} \label{SIfig:braintransfer_BA}
\end{figure}

In summary, the data-model agreement (as quantified by the modeling NRMSE) resulting from NBMF's self-supervised training is significantly improved compared to the dictionary-based supervised learning approach. One possible explanation for this phenomenon is that the \textit{model gap} between the assumed and the actual physics is the bottleneck impediment for quantitative reconstruction rather than the SNR alone. To address that, augmentation by a simple noise addition is intrinsically insufficient. This adds to the distribution mismatch between the synthetic dictionary and the in-vivo data. The latter issue could be alleviated by generating a dictionary based on sampling from a data-driven distribution. However, the former is a fundamental gap that is harder to bridge, motivating the search for alternative approaches. The self-supervised procedure emerges as a means to provide a neural reconstructor that is, by design, directly trained to (A) encapsulate the unbiased, real-world data distribution as an implicit prior, (B) counteract the model gap, and crucially, (C) predict parameter estimates with a low modeling error. 

\section*{Supplementary Note 3. Comparing NBMF to dictionary-matching and VBMF}

In this section, we expand on quantification methods that explicitly reduce the voxelwise \textit{modeling error} (becoming simply the \textit{fitting error} in this case). These include:
\begin{itemize}
    \item Dictionary matching  - which minimizes the signal discrepancy up to the grid resolution limit.
    \item Voxelwise Bloch-McConnell Fitting (\textbf{VBMF}) - a classical purely voxelwise fitting, accelerated in this work using a GPU-based JAX realization of the numerical forward simulation.
    \item Neural Bloch-McConnell Fitting (\textbf{NBMF}) in fitting mode (applied on the first subject) - jointly minimizing the signal discrepancy across all voxels while enforcing a smooth mapping from raw data signals to quantitative parameter estimates using a neural network (main text, Fig. 1).
\end{itemize}

\begin{figure}[b!]
\centering
\includegraphics[trim={0 0 0 1.5cm},clip,width=\linewidth]{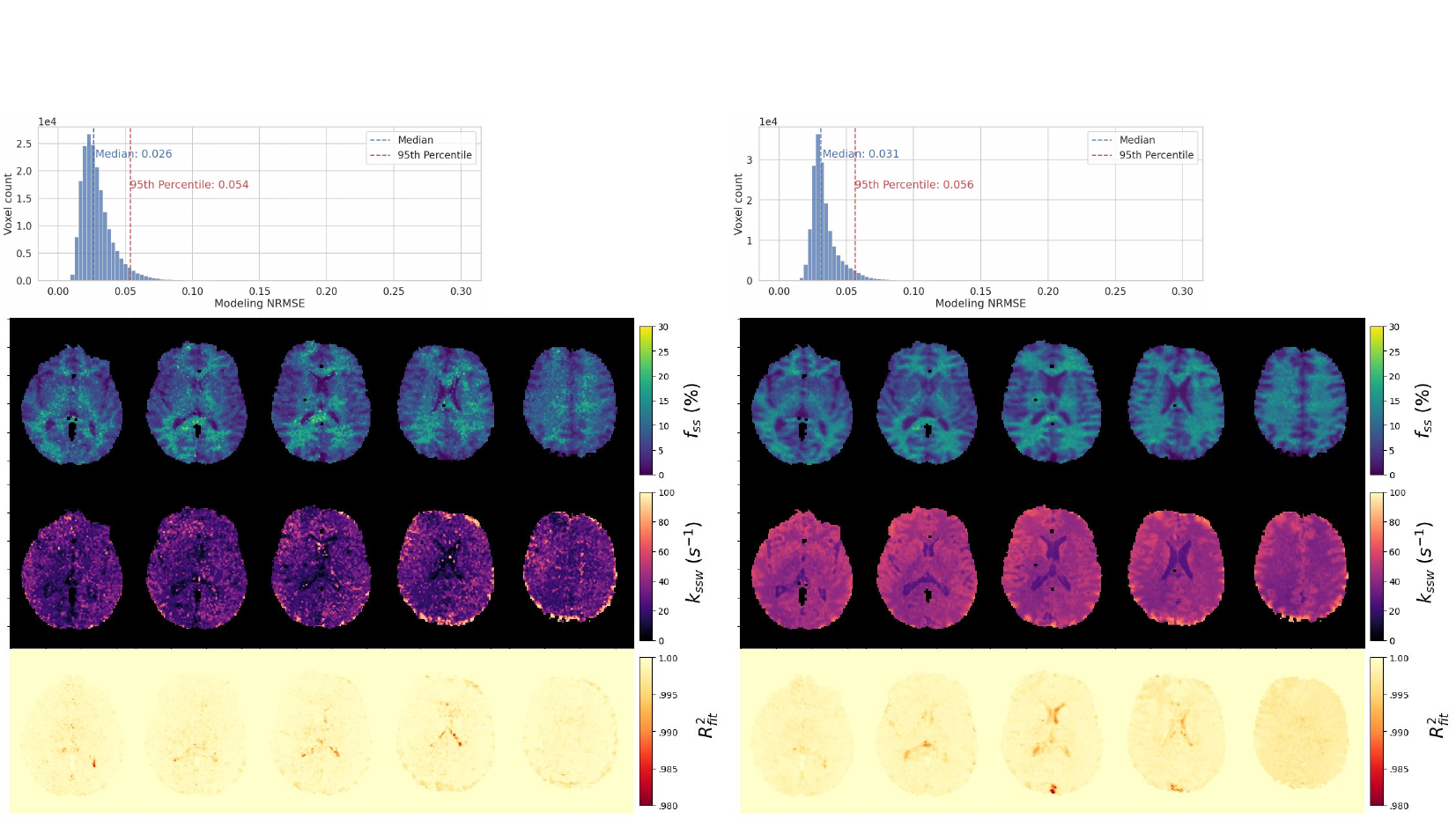}
\caption{\textbf{NBMF vs. dot-product. Bottom}: Semisolid MT-MRF reconstruction using Dictionary Matching (\textbf{left}) and NBMF (\textbf{right}), representative slices. \textbf{Top}: Corresponding distributions of the voxelwise modeling error.
} \label{SIfig:dictmatch_vs_NBMF}
\end{figure}

As a part of the dictionary-based comparative benchmark, we also implemented a standard MRF dictionary matching procedure (via dot-product)\cite{Cohen2018CESTMRF} using the GPU-accelerated JAX implementation developed in this work). The matching of the $\approx$194K voxelwise signals to the dictionary as described in the previous section (400K entries) took 93 seconds. However, the resulting parameter maps (Supplementary Fig. \ref{SIfig:dictmatch_vs_NBMF}, left) are noisy, especially for k$_{ssw}$ estimation. The use of NBMF, on the other hand, exploits information from \textit{multiple} voxels for training the reconstruction neural network, offering a joint parameter estimation that could more smoothly characterize the different tissue properties (Supplementary Fig. \ref{SIfig:dictmatch_vs_NBMF}, right).

\begin{figure}[b!]
\centering
\includegraphics[trim={0 0 0 1.5cm},clip,width=\linewidth]{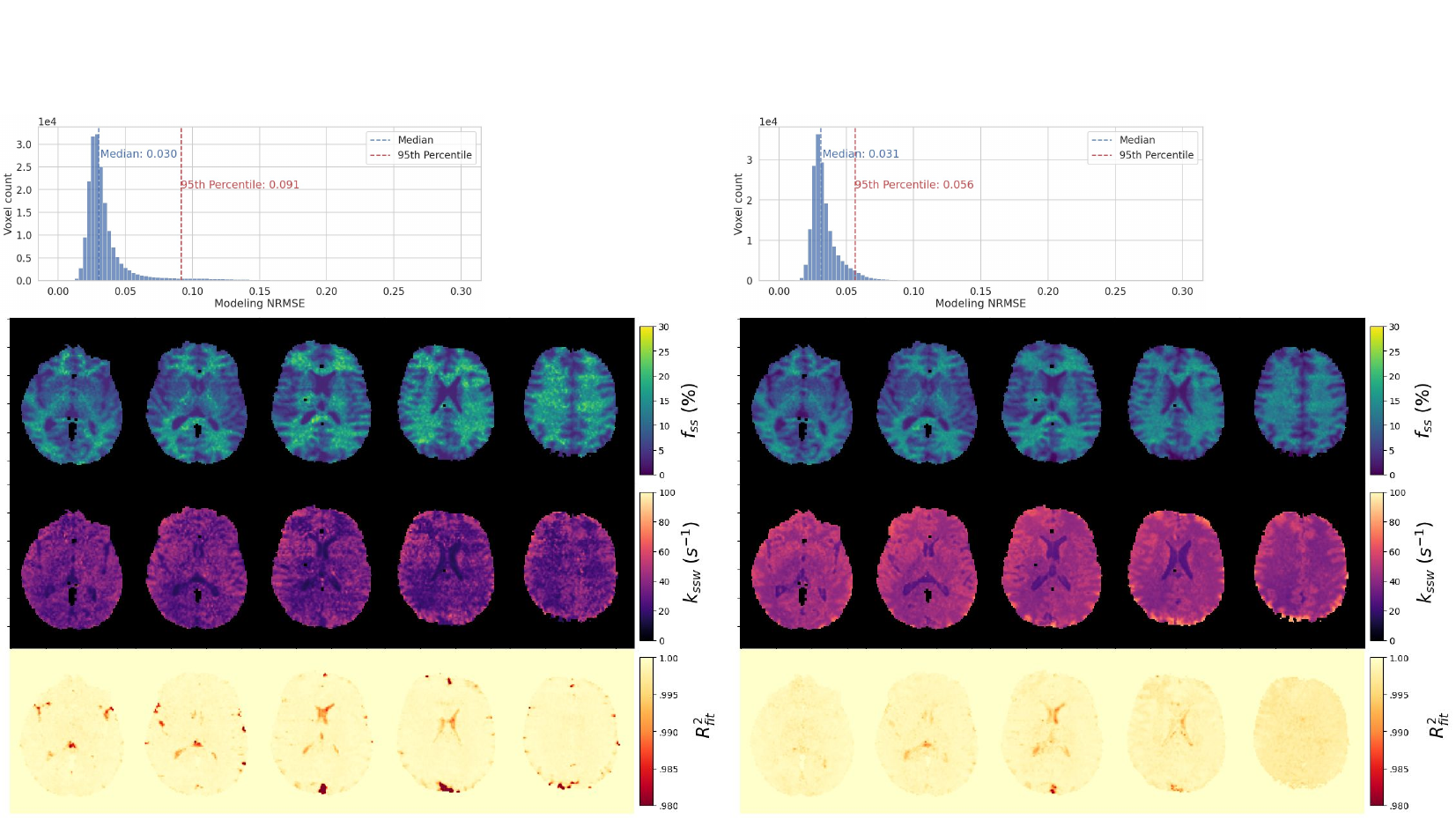}
\caption{\textbf{NBMF vs. VBMF. \ Bottom}: Semisolid MT-MRF reconstruction using VBMF (\textbf{left}) and NBMF (\textbf{right}), representative slices. \textbf{Top}: Corresponding distributions of the voxelwise modeling error.}
\label{SIfig:SI_VBMF_vs_NBMF}
\end{figure}

Next, we expand on the gradient-based voxelwise Bloch-McConnell fitting (\textbf{VBMF}), a natural simplification of the \textbf{NBMF} pipeline, achieved by stripping the neural component and stopping the gradient at the voxelwise tissue parameters which thereby become the trainable "variables" (see Fig. 1 and Methods section of the main text). Thus, the VBMF is directly comparable to legacy Bloch fitting of multi-B$_1$ Z-spectra (or CEST-MRF acquired data, given the parallel $Z_{n-1} \longrightarrow Z_n$ mode, see Supplementary Note 4). Note that as a first-order optimizer, its convergence is theoretically inferior to second-order options such as Levenberg-Marquardt (LM) which is considered standard (e.g., \textit{BM\_sim\_fit}\cite{Zaiss2018_QUESPrev} employs LM implicitly through \textsc{matlab}'s \textsc{lsqcurvefit} defaults). In practice, however, the machine-analytic GPU-accelerated solution ensures very fast convergence, as demonstrated in Supplementary Note 5. For in-vitro CEST-MRF data, NBMF and VBMF also perform similarly well, yielding results in the vicinity of the ground truth and dictionary matching results, as analyzed in Supplementary Note 1. For the in-vivo case, however, the results deteriorate owing to the higher model gap and less information from an individual voxel data. Supplementary Fig. \ref{SIfig:SI_VBMF_vs_NBMF} compares tissue parameter maps reconstructed with NBMF and VBMF from a semisolid MT MRF data acquired in a healthy human volunteer. Similar results are obtained for the estimates of the concentrations, with a denoising effect for the exchange rate observed for NBMF. 

While neighboring voxels can have significantly different fit results by traditional BM fitting, NBMF, on the other hand, fits a single model shared among the pixels, which improves data utilization and robustness to signal deviations. Notably, these benefits occur despite the network architecture being a trivial voxelwise MLP, which does not learn the brain anatomy. We predict that incorporation of spatial learning via the use of convolutional neural networks (CNNs), potentially combined with training on multiple subjects, will constitute a promising future direction for research. 

The VBMF might be a viable accelerated alternative to existing fitting methods, broadly applicable to generalized Z-spectra/CEST-MRF in-vitro experiments. Like other direct fitting methods, VBMF offers conceptual simplicity and guarantees the best fit or minimized discrepancy per voxel. This characteristic, shared with dictionary-matching methods (up to the precision limits imposed by grid quantization), is not retained in faster neural-inference methods. 

However, NBMF seems preferable for human brain imaging. As emphasized in the main text, it is able to provide a neural network (NN) reconstructor enabling real-time inference on new subjects at no extra cost. This is because the highly optimized GPU implementation makes the added computational burden of the small NN negligible. In addition to providing smoother maps, NBMF reaches convergence faster, possibly within just a few epochs, because the data redundancy relieves the model from the need to see each individual voxel many times. This effect of having the weights almost converged already when arriving at a later voxel, can be interpreted as "intra-subject transfer", by analogy to the "inter-subject" NN transfer demonstrated in the main text. The resultant advantage is of course not available for any fully voxelwise fitting method. Finally and most crucially, as shown by the comparisons (top panel in Supplementary Fig. \ref{SIfig:dictmatch_vs_NBMF}, and \ref{SIfig:SI_VBMF_vs_NBMF}), NBMF can maintain a modeling error that is close to the minimal attainable for each pixel (as attainable by dictionary-matching or VBMF). This suggests that the smoother quantitative parameter maps obtained by NBMF are similarly reliable and consistent with the raw data. {The deviation of NBMF from pure voxelwise discrepancy-minimization can be interpreted as a "biased" estimation, that is, with respect to a Maximum-Likelihood under an assumption of a physical model with additive Gaussian noise. The enforcement of a neural mapping reflects an implicit regularizing prior leading to Maximum A-Posteriori (MAP) estimation\cite{xu_test-time_2024, haltmeier_regularization_2020}. Such a data-driven prior encapsulates information from other voxels, including aggregated insight on the distribution and the model gap.

\section*{Supplementary Note 4. Extending NBMF for Z-spectra parameter fitting}
The main focus of the manuscript is on fitting non-steady-state pulsed-wave (PW) CEST-MRF data. However, both CEST-MRF and Z-spectra can be unified and represented under a generalized CEST protocol\cite{ Zaiss2018_QUESPrev}. In both cases the same blocks (readout, recovery, and saturation) are applied, with the saturation pulse power and/or frequency offset varying between different images. The recently adopted accelerated Z-spectra collection\cite{Zaiss2018_QUESPrev} is also non-steady-state, with larger but still finite recovery times (e.g., $T_{rec}$ = 3 sec); the effect is approximated\cite{ Zaiss2018_QUESPrev} by a history-independent constant factor (i.e., assume saturation and acquisition bring z-magnetization to zero, so the pre-saturation magnetization obeys: $M_{ini}\approx M_0 e^{-T_{rec}}/T_1$ , e.g., $Z_{ini}\approx$0.95 for $T_1\approx$1 sec). In Z-spectra modeling this is typically joined by other approximations for the complicating effects of multiple pools, pulsed and finite (partial saturation) RF irradiation. Here, we analyze Z-spectra sequences as an instance of CEST-MRF, with the saturation parameters following a specific sequential pattern, without the above approximations for the dynamics. 

The insight we offer here is that the initial condition of each experiment in each voxel is known (to the precision permitted by the measurement) by virtue of the previous readout. This removes the need for a sequential simulation of the entire magnetization history. The initial (normalized) pre-saturation magnetization $Z_i$ can be estimated for each iteration by processing the result of the "snapshot"\cite{Mueller2020} acquisition through simulation of all flips and relaxations throughout the readout and recovery stages as follows: 
\begin{equation}
\hspace{120pt}  
\Tilde{Z}_{ini}[n] = 1 - (1 - \mathsf{R}(Z_{acq}[n-1]))e^{-T_{rec}/T_1}
\label{eq_relax}    
\end{equation}
where $\mathsf{R}$ represents the operator of (normalized) Z-magnetization evolution across the readout stage involving multiple small-flip-angle rotation and pause blocks, with spoilers applied.
It is assumed the acquired decoded signal can be translated to Z-value via normalizing voxel-wise by the equilibrium (no-saturation) signal reference, e.g. the first B$_1$[n=0]=0 iteration. The prediction for the next acquisition will thus be:
\begin{equation}
\hspace{120pt}
    \Tilde{Z}_{acq}[n]=\mathsf{S}_n(Z_{ini}[n])=\mathcal{F}_n(Z_{acq}[n-1])
    \label{SIeq_BM}  
\end{equation} where $\mathsf{S}$ is a Bloch-McConnell simulator of the saturation stage, $\mathcal{F}_n$ is the overall evolution across one iteration, including readout ($\mathsf{R}$), recovery (as formulated explicitly in Supplementary Equation \eqref{eq_relax}) and saturation ($\mathsf{S}_n$) stages, and $Z_{acq}[n=0]=1$. 

This enables us to avoid sequential modeling of the hundreds of iterations that constitute the multi-B1 Z-spectra collection, and to achieve parallel processing for both forward simulation and inverse fitting. We thus treat both types of experiments the same, by ignoring the Z-spectra special sequence and viewing it just like CEST MRF: a batch of magnetization evolution instances with known initial and final conditions, varying saturation RF parameters and common unknown tissue parameters. These values can be obtained by fitting, i.e. seeking best reconstruction of final (normalized) magnetizations $\{Z_{acq}[n]\}_{n=1}^N$ from initial ones $\{Z_{ini}[n]\}_{n=1}^N$ under the model (Supplementary Equation \eqref{SIeq_BM}), computed as the regression: $  \{f,k\} = argmin_{\{f,k\}} \sum_n||\mathcal{F}(Z_{n-1})-Z_{n}||$. 

Within the technical context of the NBMF framework, the above analysis leads to two options: sequential or parallel execution of the simulation. The parallel option results in faster training in the full numerical (3-pool) case, especially for Z-spectra, which involve an order-of-magnitude longer sequence. We therefore use this option for applications of NBMF to quantification of parameters from either CEST-MRF (see main text) or multi-B$_1$ Z-spectra  (see next section).

More generally, the insight regarding the ability to use non-constant $Z_i$ computed from the previous iteration can be applied to various existing analytical or numerical methods for fitting Z-spectra. Furthermore, it can inform acquisition design by suggesting reduced or possibly non-uniform recovery times.

\section*{Supplementary Note 5.  Z-spectrum fitting speed benchmark}

The speed of the proposed approach was compared to traditional numerical BM-fitting\cite{Zaiss2018_QUESPrev, zaiss2013chemical}, as implemented using a common CEST MATLAB repository\cite{Zaiss2018_QUESPrev} (\url{https://github.com/cest-sources/BM_sim_fit}, commit 029e00c). The same desktop workstation, equipped with 24 Intel i9 CPUs and a single Nvidia GeForce RTX 3060 GPU was used in both cases. The benchmark data were obtained by simulating synthetic multi-B$_1$ Z-spectra\cite{Zaiss2018_QUESPrev}, assuming three proton pools: (a) water, (b) amide, and (c) semi-solid (MT) with parameters oriented for white matter (T$_1$=1.5s, T$_{2w}$=49ms, T$_{2s}$=40ms, T$_{2ss}$=40$\mu$s, f$_{s}$=0.2\%, k$_{sw}$=300$s^{-1}$, f$_{ss}$=20\%, k$_{ssw}$=70$s^{-1}$). The field strength was set to B$_0$=7T, the saturation pulse powers were B$_1$=[.6, .9, 1.5, 2.0] $\mu$T, and the saturation pulse frequency offset was varied non-uniformly between -30 and 30 ppm, with the majority of samples concentrated between -5 and 5 ppm in 0.25 ppm increments, for a total of 61 samples per B$_1$ value and 244 overall. A 1\% noise was added to the signal.  The full sequence details are available in: \url{https://github.com/cest-sources/MultiB0_B1_qCEST_brain}. 
An excellent fit ($R^2$=99.8) was obtained by both methods (see Supplementary Fig. \ref{SIfig:synthZspec_fit}). The process was repeated using different starting values for the proton volume fraction and exchange rate estimates, demonstrating similar convergence to the correct values in all cases (Supplementary Fig. \ref{SIfig:synthZspec_fit}c,d), even for the more challenging k$_{sw}$ estimate.

The speed performance of both methods is presented in the
Supplementary Table \ref{tab:runtimes}. Traditional BM fitting requires ~40 seconds per voxel, with the overall time scaling linearly with the number of voxels to reach a total reconstruction time of 160 hours, impractical for 3D human imaging. Initializing NBMF processing in the JAX framework\cite{jax2018github} requires roughly a minute for graph compilation and GPU allocation, but this is quickly amortized by the speed of subsequent highly optimized rapid fitting iterations. For a medium-sized single brain image slice of 128x128$\sim$16K voxels, the runtime for NBMF is 3 orders of magnitude shorter than the benchmark.  


\begin{table}[ht]
\centering
\begin{tabulary}{\textwidth}{|L|C|R|}
    \hline
    \\ Numerical fitting method & \textbf{NBMF} (this work) & \textbf{BM fitting}\cite{ Zaiss2018_QUESPrev}  
    \\ \hline
    10 voxels reconstruction time & 30 sec  (measured) & 400 sec (measured) \\ \hline
    16K voxels reconstruction time & 200 sec (measured) & 160 hours (estimated) \\ \hline
\end{tabulary}
\caption{\textbf{NBMF yields $\times$2880 acceleration over BM fitting.}}
\label{tab:runtimes}
\end{table}

\begin{figure}[ht]
\centering
\includegraphics[trim={0 0 0 0.3cm},clip,width=0.95\linewidth]{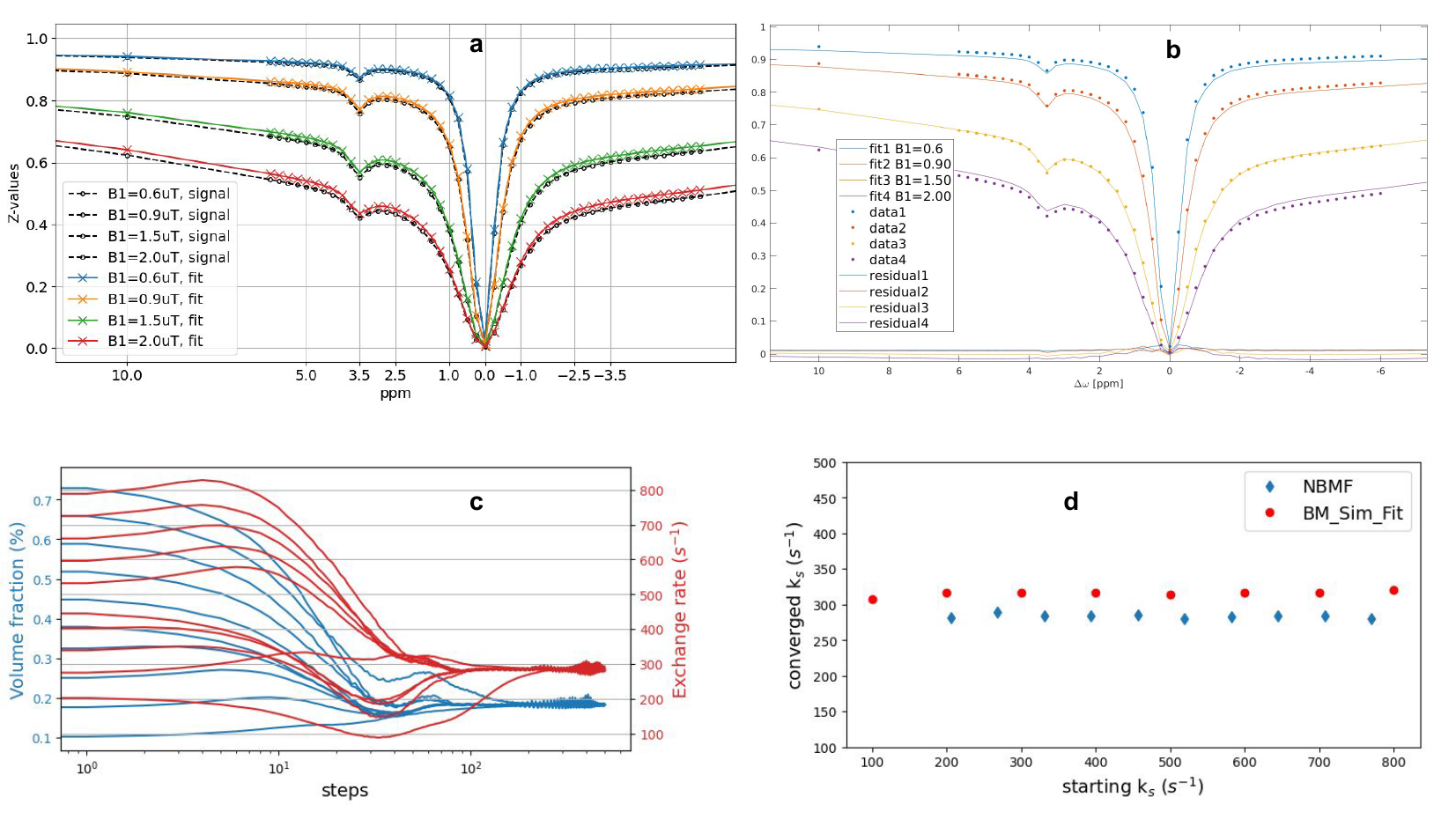}
\caption{\textbf{NBMF applied to classical Z-spectra}. Fitting results for a synthetic 3-pool multi-B$_1$ Z-spectra data for NBMF (a) and standard BM fitting (BM\_sim\_fit routine) (b). The process was repeated using different initialization of the parameter estimates (c, d), yielding a consistent and effective convergence in both cases.}
\label{SIfig:synthZspec_fit}
\end{figure}

\section*{Supplementary Note 6. NBMF results for the estimation of quantitative exchange parameters from semisolid MT-MRF and CEST-MRF data of n = 4 healthy volunteers}

\begin{figure}[H]
\centering
\includegraphics[trim={0 0 0 5cm},clip, height=0.9\textheight]{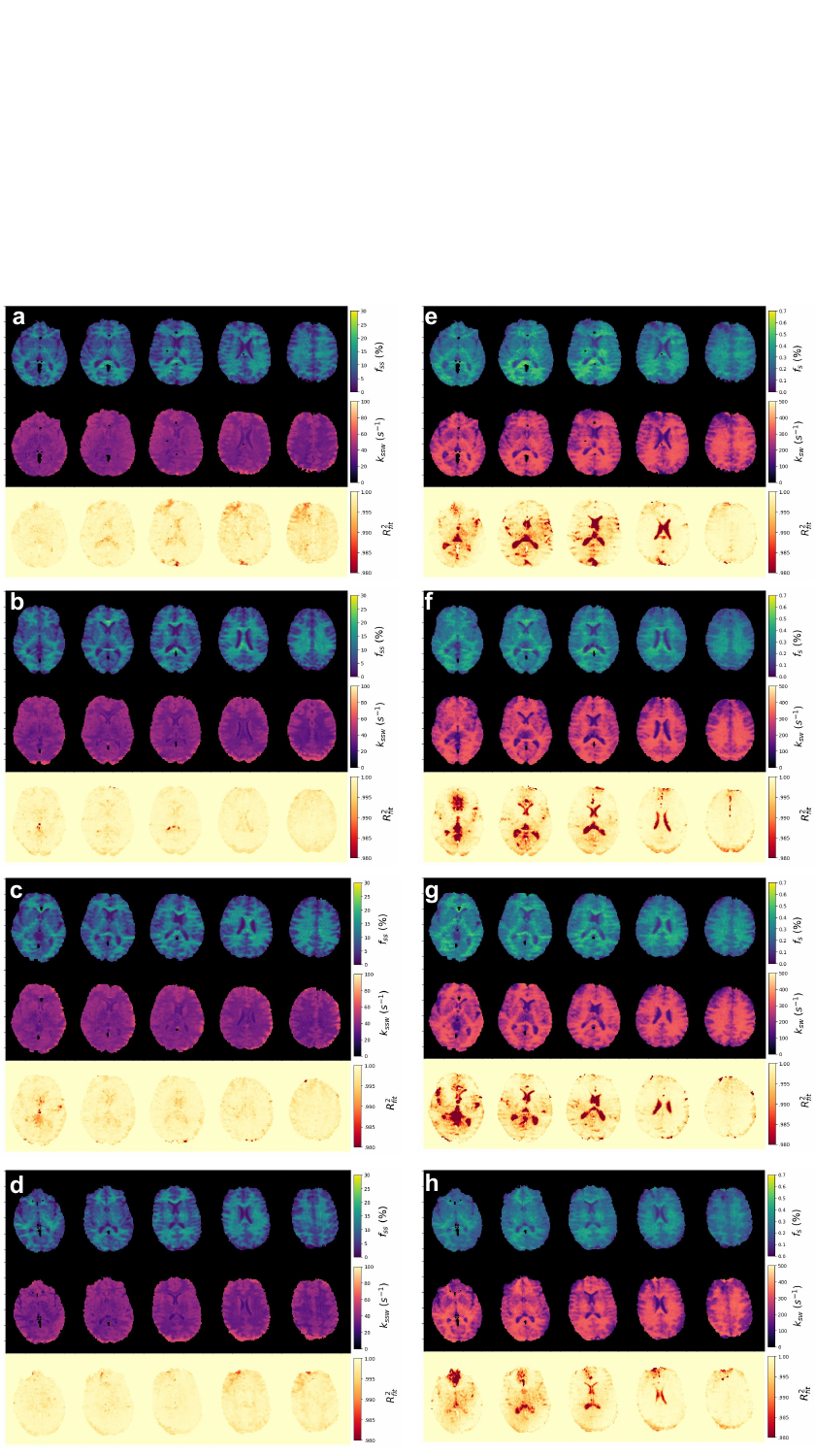}
\caption{Tissue parameter quantitative maps obtained by applying NBMF-trained rapid neural reconstructors for representative slices of the 4 healthy volunteers, for semisolid-MT pool (a-d) and Amide pool (e-h).}
\label{fig:all_brains_nbmf}
\end{figure}

\flushbottom
\section*{Supplementary Note 7. Auxiliary data overview}

\begin{figure}[H]
\centering
\includegraphics[trim={0 0 0 8cm},clip, height=0.9\textheight]{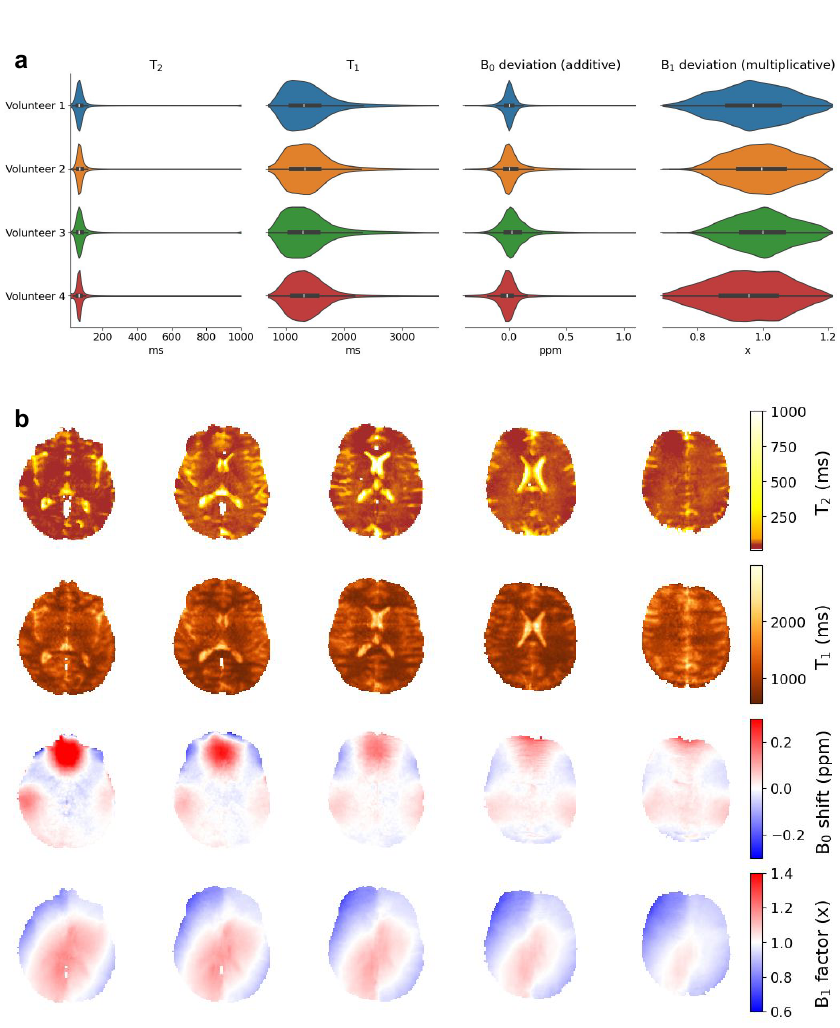}
\caption{\textbf{Overview of T$_1$, T$_2$, B$_0$, B$_1$ data}. (a) Parameter value distribution across the entire brain data from n=4 healthy volunteers. (b) Representative auxiliary parameter maps from volunteer \#1.}
\label{fig:auxiliart}
\end{figure}
\vfill

\bibliography{mendeley, aux}

\end{document}